\documentclass[]{youtu} % For LaTeX2e

\usepackage{mathpazo}
\usepackage{graphicx}
\usepackage[numbers]{natbib}
\usepackage{float}
\usepackage{enumitem}
\usepackage{graphicx}
\usepackage[most]{tcolorbox}
\usepackage{subcaption}

\usepackage{tcolorbox}
\usepackage{enumitem}
\usepackage[dvipsnames]{xcolor}
\usepackage{fontawesome}
\usepackage{multirow}

\usepackage[utf8]{inputenc} % allow utf-8 input
\usepackage[T1]{fontenc}    % use 8-bit T1 fonts
\usepackage{hyperref}       % hyperlinks
\usepackage{url}            % simple URL typesetting
\usepackage{booktabs}       % professional-quality tables
\usepackage{amsfonts}       % blackboard math symbols
\usepackage{nicefrac}       % compact symbols for 1/2, etc.
\usepackage{microtype}      % microtypography
\usepackage{xcolor}         % colors
\usepackage{amsmath}
\usepackage{enumitem}
\usepackage[most]{tcolorbox}
\usepackage{makecell}

\usepackage{graphicx}
\usepackage{multirow}
\usepackage{colortbl}
\usepackage[table]{xcolor}
\usepackage{array}
\usepackage{makecell}
\usepackage{makecell}
\usepackage{arydshln} % 支持虚线
\usepackage{wrapfig}
\usepackage{algorithm}
\usepackage{algpseudocode}
\usepackage{pgfplots} 
\usepackage{tabularx}
\makeatletter
\renewcommand\tableofcontents{%
  \@starttoc{toc}%
}
\let\origaddcontentsline\addcontentsline
\newcommand{\DisableTOC}{\let\addcontentsline\@gobblethree}
\newcommand{\EnableTOC}{\let\addcontentsline\origaddcontentsline}
\makeatother

\definecolor{myborder}{RGB}{73, 86, 102}
\definecolor{myRed}{RGB}{240, 48, 159}
\definecolor{mylightblue}{RGB}{235, 245, 255}

\tcbuselibrary{skins}

\title{Beyond the All-in-One Agent: Benchmarking
Role-Specialized Multi-Agent Collaboration in
Enterprise Workflows}
\author{
Tao Yu\textsuperscript{$1,2,3 * \S$}, 
Hao Wang\textsuperscript{$2, * \dag$},  
Changyu Li\textsuperscript{$1,4 \S$}, 
Shenghua Chai\textsuperscript{$2 \dag$}, 
Minghui Zhang\textsuperscript{$2 \dag$}, \\
Zhongtian Luo\textsuperscript{$2 \dag$}, 
Yuxuan Zhou\textsuperscript{$5$}, 
Haopeng Jin\textsuperscript{$2 \dag$}, 
Zhaolu Kang\textsuperscript{$4$},, 
Jiabing Yang\textsuperscript{$2,3$}, \\
YiFan Zhang\textsuperscript{$2,3$}, 
Xinming Wang\textsuperscript{$2,3$}, 
Hongzhu Yi\textsuperscript{$3$}, \\
Zheqi He\textsuperscript{$1 \ddagger$}, 
JingShu Zheng\textsuperscript{$1$}, 
Xi Yang\textsuperscript{$1$}, 
Yan Huang\textsuperscript{$2,3 \ddagger$}, 
Liang Wang\textsuperscript{$2,3$}
}

\vspace{-5mm}

\affiliation{
\textsuperscript{$1$}BAAI
\textsuperscript{$2$}CASIA \ 
\textsuperscript{$3$}UCAS \ 
\textsuperscript{$4$}Peking University \
\textsuperscript{$5$}Tsinghua University
}

\date{May 9, 2026}
\sourcecode{https://github.com/yutao1024/EntCollabBench}
\data{https://huggingface.co/datasets/Kirito-Lab/EntCollabBench}
\begin{document}

\DisableTOC

\abstract{Large language model (LLM) agents are increasingly expected to operate in enterprise environments, where work is distributed across specialized roles, permission-controlled systems, and cross-departmental procedures. However, existing enterprise benchmarks largely evaluate single agents with broad tool access, while existing multi-agent benchmarks rarely capture realistic enterprise constraints such as role specialization, access control, stateful business systems, and policy-based approvals. We introduce \textsc{EntCollabBench}, a benchmark for evaluating enterprise multi-agent collaboration. \textsc{EntCollabBench} simulates a permission-isolated organization with 11 role-specialized agents across six departments and contains two evaluation subsets: a Workflow subset, where agents collaboratively modify enterprise system states, and an Approval subset, where agents make policy-grounded decisions. Evaluation is based on execution traces, database state verification, and deterministic policy adjudication rather than natural-language response judging. Experiments with representative LLM agents show that current models still struggle with end-to-end enterprise collaboration, especially in delegation, context transfer, parameter grounding, workflow closure, and decision commitment. \textsc{EntCollabBench} provides a reproducible testbed for measuring and improving agent systems intended for realistic organizational environments.}
\maketitle

\renewcommand{\thefootnote}{*}
% \footnotemark % creates the * on the page where you want it  
\footnotetext{Equal contribution.}

\renewcommand{\thefootnote}{\dag}
\footnotetext{Work done during an internship at CASIA.}

\renewcommand{\thefootnote}{\S}
\footnotetext[0]{Work done during an internship at BAAI.}

\renewcommand{\thefootnote}{\textdaggerdbl}
\footnotetext{Corresponding author.}

% \renewcommand{\thefootnote}{\ensuremath{\spadesuit}}
% \footnotetext{Project leader.}
\renewcommand{\thefootnote}{\arabic{footnote}}

\vspace{-.1em}

\section{Introduction}

Large language model (LLM) agents~\citep{yu2025browseragentbuildingwebagents} are increasingly expected to operate in enterprise environments~\citep{agarwal2026enterpriselabfullstackplatformdeveloping,liu2025agentbenchevaluatingllmsagents,drouin2024workarenacapablewebagents,boisvert2025workarenacompositionalplanningreasoningbased,xu2025theagentcompanybenchmarkingllmagents}, where work is distributed across specialized roles, permission-controlled systems, and cross-departmental procedures. Completing a business request therefore requires more than isolated tool use: agents must infer responsibility boundaries, delegate to the right role, preserve context, and update stateful systems correctly.

Recent enterprise agent benchmarks have advanced realistic workplace evaluation by using service platforms, CRM systems, code repositories, and collaboration tools~\citep{malay2026enterpriseopsgymenvironmentsevaluationsstateful}. However, most still assume a single agent with broad tool access, which differs from real organizations where HR, IT, customer support, engineering, and approval roles have separate responsibilities and permissions. Conversely, existing multi-agent benchmarks study communication and coordination, but usually in games, puzzles, or abstract distributed-information settings rather than enterprise workflows with access control, persistent records, and policy constraints~\citep{samvelyan2019starcraftmultiagentchallenge,ruhdorfer2025overcookedgeneralisationchallengeevaluating,ossowski2025commacommunicativemultimodalmultiagent}.

We introduce \textsc{EntCollabBench}, a benchmark for evaluating enterprise multi-agent collaboration. It simulates a permission-isolated organization with 11 role-specialized agents across six departments: IT, Human Resources, Customer Service, Shared Services, Engineering, and an Approval Center. Given a natural-language instruction and a designated starting agent, agents must complete tasks through cross-departmental delegation and communication, using only tools within their assigned responsibility scope.

\textsc{EntCollabBench} contains two evaluation subsets. The \textbf{Workflow subset} covers operational tasks that modify enterprise system state, such as creating incidents, updating HR cases, scheduling meetings, revising knowledge articles, and submitting pull requests. The \textbf{Approval subset} covers policy-grounded decisions by finance, legal, and procurement specialists. Workflow tasks are evaluated through execution traces and database state diffs, while approval tasks are evaluated against deterministic policy adjudications.

Experiments show that current LLM agents still struggle with enterprise collaboration. Although many models can perform local role-specific actions, end-to-end success drops markedly when tasks require multi-hop delegation, context transfer, final-stage coordination, and parameter-level grounding in stateful systems.

Our contributions are:

We formulate enterprise multi-agent collaboration as an evaluation setting centered on role specialization, permission isolation, implicit routing, context transfer, and stateful cross-departmental workflows.

We introduce \textsc{EntCollabBench}, a benchmark with 11 role-specialized agents across six departments, covering both operational workflow execution and policy-based approval decisions with objective verification.

We evaluate representative LLM agents on \textsc{EntCollabBench} and identify key bottlenecks in delegation, parameter grounding, workflow closure, decision commitment, and coordination cost.

\begin{figure}
    \centering
    \includegraphics[width=0.95\linewidth]{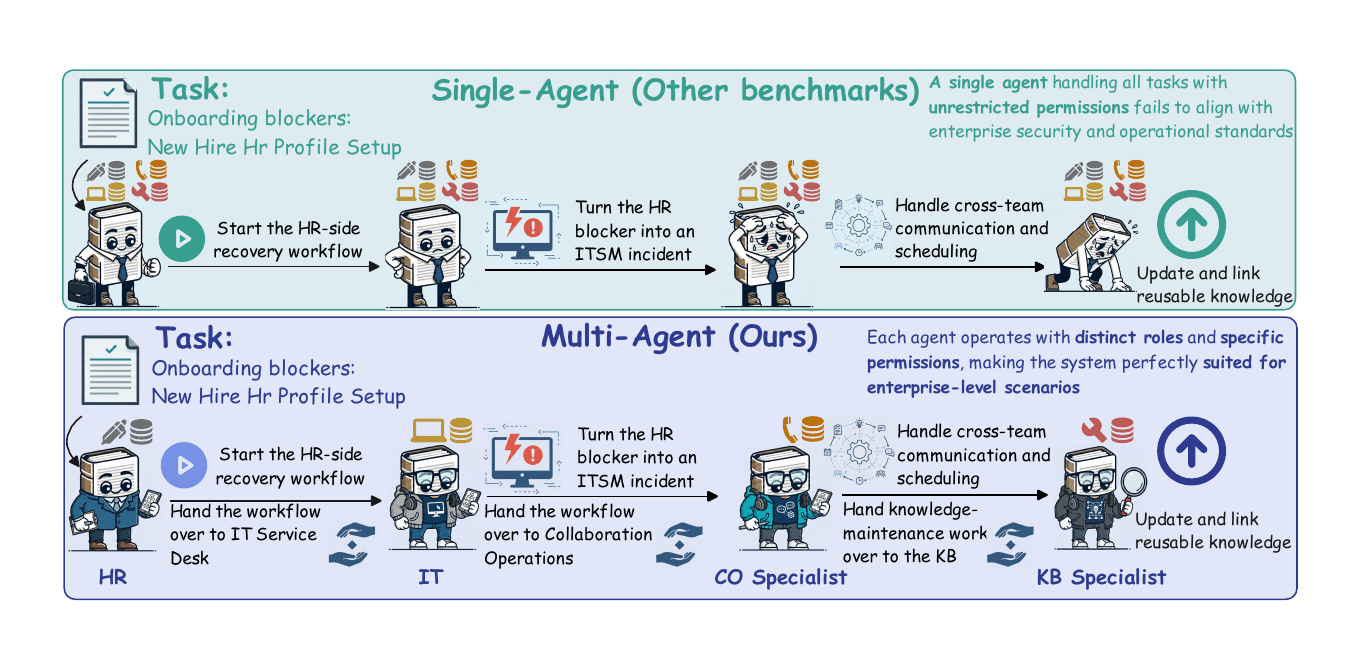}
    \vspace{-5pt}
    \caption{Comparison of EntCollabBench with other enterprise benchmarks.}
    \label{1}
    \vspace{-13pt}
\end{figure}

% \section{Related Works}

% 单智能体企业Benchmark。 近年来，面向企业场景的智能体评测基准发展迅速。AgentBench [1] 首次系统评估了LLM作为智能体在多种环境中的表现。WorkArena [2] 和 WorkArena++ [3] 基于ServiceNow平台构建了面向知识工作者的Web任务集。TheAgentCompany [4] 模拟了包含GitLab、RocketChat等工具的公司环境，评估智能体完成真实企业任务的能力。EntWorld [5] 进一步扩展至CRM、ITSM等六个企业领域的1756个GUI任务。AgentArch [6] 评估了18种智能体架构配置在企业工作流上的表现。此外，Finch [7] 和 CRMArena [8] 分别针对财务会计和CRM场景构建了领域专用基准。然而，上述工作均采用单智能体范式——由一个智能体承担所有角色的任务，未涉及智能体间的跨角色通信与协作，与真实企业中多部门分工协作的工作模式存在显著差距。

% 多智能体协作Benchmark。 在多智能体评测方面，早期工作集中于强化学习领域，如SMAC [9] 和 Overcooked-AI [10] 评估了游戏环境中的合作策略。进入LLM时代后，COMMA [11] 评估了信息不对称条件下多模态智能体的通信协作能力，MultiAgentBench [12] 在Minecraft等场景中衡量了合作与竞争行为，SILO-BENCH [13] 测试了分布式信息条件下的多智能体协调能力。这些工作虽然涉及智能体间通信，但场景均为游戏、拼图或算法任务，未涉及更加复杂真实的场景，如企业工作流。

% 基于角色分工的多智能体系统。 ChatDev [14] 和 MetaGPT [15] 引入了CEO、程序员、测试员等角色分工，通过结构化通信完成软件开发任务。Generative Agents [16] 在虚拟小镇中模拟了具有不同身份的25个智能体的社会交互。ClinicalLab [17] 在医疗领域实现了跨科室的多智能体诊断协作。OrgAgent [18] 提出了类公司组织结构的多智能体框架。然而，这些工作要么局限于软件开发单一领域，要么是框架而非评测基准，均未构建面向通用企业场景的多智能体协作评测任务集。

\section{Related Works}

\textbf{Single-Agent Enterprise Benchmarks.}
In recent years, agent evaluation benchmarks targeting enterprise scenarios have advanced rapidly. AgentBench~\citep{liu2025agentbenchevaluatingllmsagents} was the first to systematically evaluate LLMs as 
agents across diverse environments. WorkArena~\citep{drouin2024workarenacapablewebagents} and WorkArena++~\citep{boisvert2025workarenacompositionalplanningreasoningbased} built web-based task suites for knowledge workers on the ServiceNow platform. TheAgentCompany~\citep{xu2025theagentcompanybenchmarkingllmagents} simulated a corporate environment equipped with tools such as GitLab and RocketChat to assess agents on realistic enterprise tasks. EntWorld~\citep{mo2026entworldholisticenvironmentbenchmark} further scaled to 1,756 GUI tasks spanning six enterprise domains including CRM and ITSM. In addition, Finch~\citep{dong2026finchbenchmarkingfinance} and CRMArena~\citep{huang2025crmarenaunderstandingcapacityllm} constructed domain-specific benchmarks for finance \& accounting and CRM, respectively. However, all of the above adopt a single-agent paradigm---one agent assumes every role---without involving inter-agent communication or cross-role collaboration, leaving a significant gap with the multi-departmental division of labor found in real enterprises. The differences are shown in Figure \ref{1}.

\textbf{Multi-Agent Collaboration Benchmarks.}
Early work on multi-agent evaluation centered on reinforcement learning, where SMAC~\citep{samvelyan2019starcraftmultiagentchallenge} and Overcooked-AI~\citep{ruhdorfer2025overcookedgeneralisationchallengeevaluating} assessed cooperative strategies in game environments. In the LLM era, COMMA~\citep{ossowski2025commacommunicativemultimodalmultiagent} evaluated communicative collaboration among multimodal agents under information asymmetry, MultiAgentBench~\citep{zhu2025multiagentbenchevaluatingcollaborationcompetition} measured cooperative and competitive behaviors in scenarios such as Minecraft, and SILO-BENCH~\citep{zhang2026silobenchscalableenvironmentevaluating} tested multi-agent coordination under distributed information.
Although these works involve inter-agent communication, their scenarios are limited to games, puzzles, or algorithmic tasks and do not address more complex, real-world settings such as enterprise workflows.

% \paragraph{Role-Based Multi-Agent Systems.}
% ChatDev~\citep{qian2024chatdevcommunicativeagentssoftware} and MetaGPT~\citep{hong2024metagptmetaprogrammingmultiagent} introduced role specialization---CEO, programmer, tester, etc.---and completed software development tasks through structured communication. Generative Agents~\citep{park2023generativeagentsinteractivesimulacra} simulated social interactions among 25 agents with distinct identities in a virtual town. ClinicalLab~\citep{yan2024clinicallabaligningagentsmultidepartmental} realized cross-departmental multi-agent diagnostic collaboration in the medical domain. OrgAgent~\citep{wang2026orgagentorganizemultiagentlike} proposed a multi-agent framework organized like a corporate hierarchy. Nevertheless, these efforts are either confined to the single domain of software development or constitute frameworks rather than evaluation benchmarks, and none has constructed a multi-agent collaborative evaluation task suite for general enterprise scenarios.

\begin{figure}
    \centering
    \includegraphics[width=0.95\linewidth]{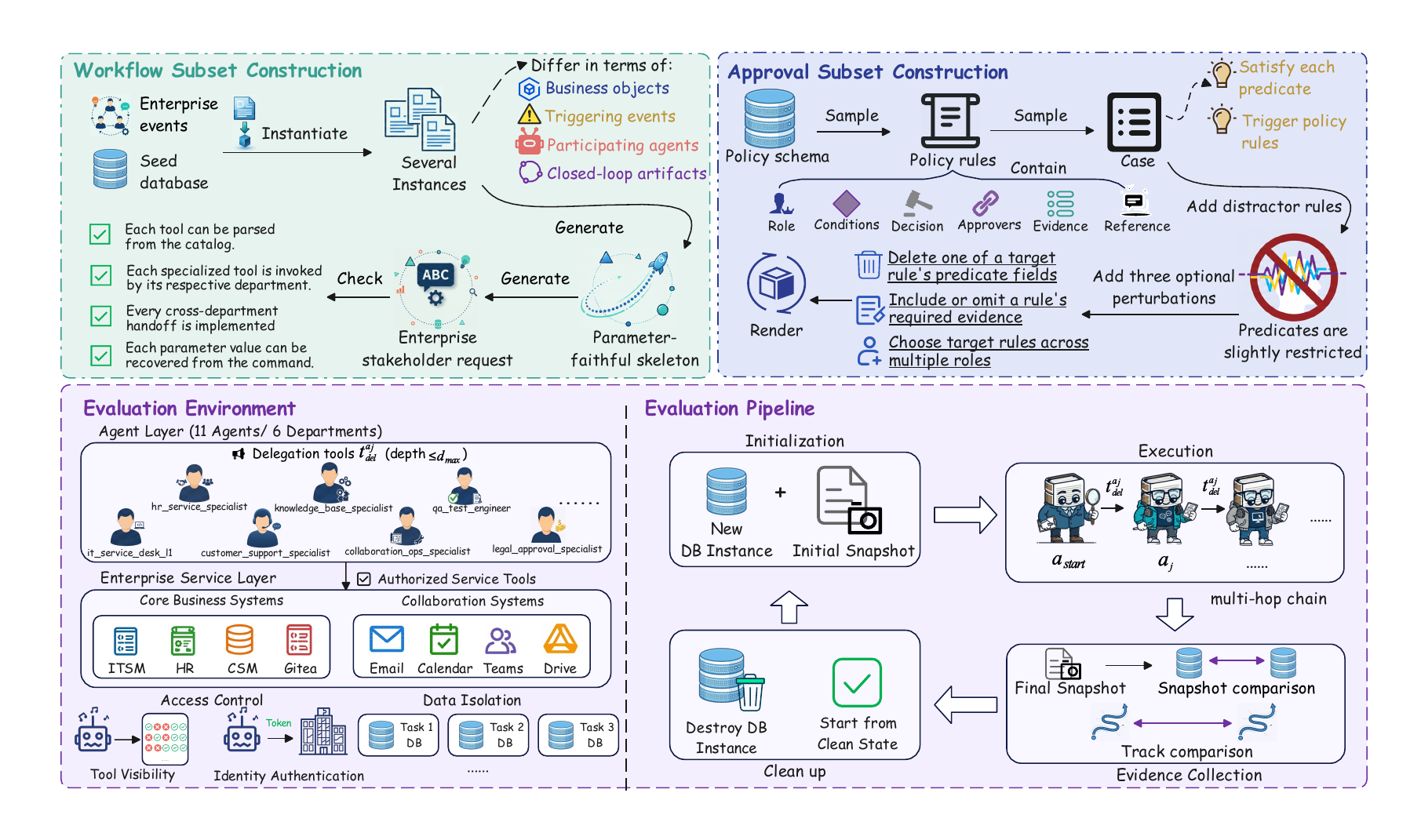}
    % \vspace{-5pt}
    \caption{Overview of EntCollabBench. The Workflow Track generates tasks across business domains and process intents, producing instances with different objects, events, agents, and artifacts. The Approval Track constructs requests from sampled rules with predicate satisfaction and optional perturbations. The Evaluation Environment includes 11 agents over 6 departments with controlled access to enterprise systems. The Evaluation Pipeline proceeds through DB initialization, multi-hop execution starting from a designated agent, snapshot and trace event collection, and DB cleanup.}
    \label{fig:2}
    \vspace{-13pt}
\end{figure}

\section{EntCollabBench}
EntCollabBench (Figure~\ref{fig:2}) evaluates whether large language model agents can perform real enterprise work under the constraints that distinguish organisational settings from sandboxed task environments: role specialization, permission isolation, stateful business systems, and cross-departmental delegation. 

\subsection{Task Definition and Design Scope}

% We propose a benchmark for enterprise multi-agent collaboration under strict permission isolation. The environment contains 11 role-specialized agents across six departments: IT, Human Resources, Customer Service, Shared Services, Engineering, and an Approval Center. Each agent can access only the tools within its responsibility scope, such as ITSM, HR, CSM, Gitea, Email, Calendar, Teams, or Drive. Given a natural-language instruction and a starting agent, agents must complete the task through cross-departmental delegation and communication.

We define each benchmark instance as a constrained collaboration task in the organizational environment introduced above. Given a natural-language instruction and a designated starting agent, agents must complete the task using only role-authorized tools and explicit cross-role delegation. Thus, success depends on responsibility inference, context transfer, and stateful workflow execution rather than isolated tool use.

This task definition induces four design constraints. First, \textbf{role isolation} is strictly enforced: agents cannot directly call tools or query data outside their own department, and must instead communicate necessary context to the appropriate downstream role. This creates information asymmetry, where insufficient delegation messages can cause execution failures, while excessive or noisy context may mislead downstream agents. Second, \textbf{collaboration dependency} is required: every task involves agents from at least two departments, ensuring that success depends on multi-agent planning and communication rather than single-agent tool use. Third, tasks require \textbf{implicit routing and decomposition}: natural-language instructions describe business goals rather than explicit workflows, so agents must infer responsibility boundaries, identify the next role, and decide which actions to perform. Finally, the workflows are \textbf{long-horizon and stateful}, involving multiple tool calls and delegation rounds, where local errors in parameters, routing, or semantics can silently propagate and cause end-to-end failure.

The task suite covers realistic enterprise operations, including ticket handling, incident response, onboarding coordination, customer escalation, knowledge-base maintenance, document approval, and code review. Each task is paired with deterministic ground truth specifying the expected final system state, key parameter constraints, or policy decision outcome. Evaluation is therefore objective and reproducible: it is based on system-state changes, critical tool-call parameters, and policy-grounded decisions rather than the agents' natural-language responses.

\subsection{Evaluation Subsets}

% \begin{wrapfigure}[26]{r}{0.48\textwidth}
% \vspace{-0.45cm}
% \centering
% \includegraphics[width=0.48\textwidth]{figs/pie.pdf}
% \vspace{-0.2cm}
% \caption{
% % Dataset statistics.
% % The benchmark contains 300 tasks in total: 
% % 160 workflow tasks, 40 workflow multi-task tasks, 
% % 80 approval tasks, and 20 approval multi-task tasks. 
% % Workflow tasks are organized into six categories, 
% % while approval tasks are organized into seven categories.
% Dataset statistics. The benchmark contains 300 tasks: 160 workflow, 40 workflow multi-task, 80 approval, and 20 approval multi-task tasks, spanning six workflow and seven approval categories.
% }
% \label{fig:3}
% % \vspace{-5.5cm}
% \end{wrapfigure}

\begin{wrapfigure}[20]{r}{0.48\textwidth}
\vspace{-0.45cm}
\centering
\includegraphics[width=0.48\textwidth]{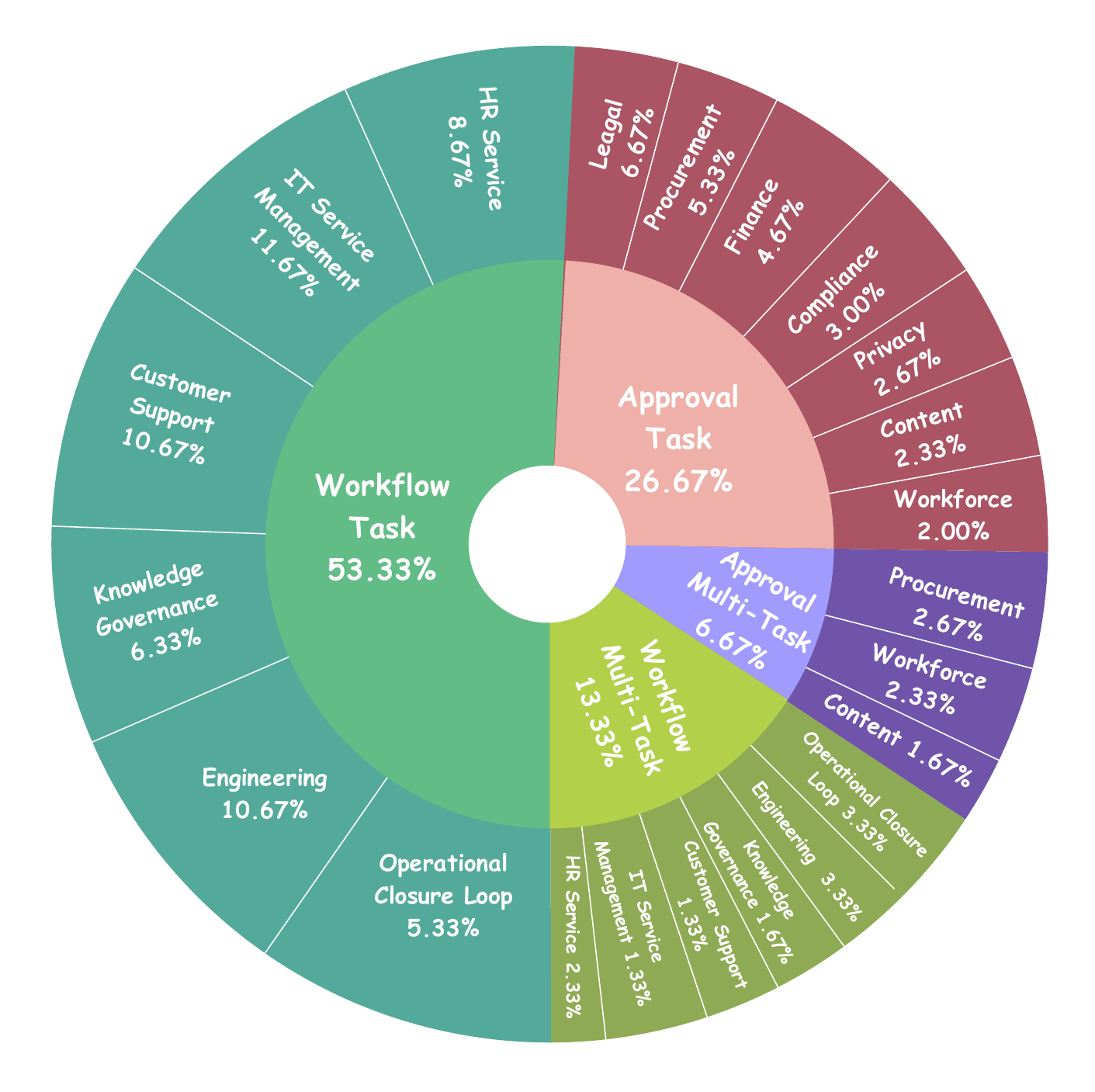}
\vspace{-0.2cm}
\caption{
% Dataset statistics.
% The benchmark contains 300 tasks in total: 
% 160 workflow tasks, 40 workflow multi-task tasks, 
% 80 approval tasks, and 20 approval multi-task tasks. 
% Workflow tasks are organized into six categories, 
% while approval tasks are organized into seven categories.
Dataset statistics. The benchmark contains 300 tasks: 160 workflow, 40 workflow multi-task, 80 approval, and 20 approval multi-task tasks, spanning six workflow and seven approval categories.
}
\label{fig:3}
% \vspace{-5.5cm}
\end{wrapfigure}

Enterprise work in our benchmark is divided into two structurally distinct subsets. The dataset statistics are shown in Figure~\ref{fig:3}.

% \textbf{Workflow subset} covers operational departments, including IT, HR, Customer Service, Shared Services, and Engineering. Tasks in this subset require agents to change external enterprise system states, such as creating incidents, updating HR cases, sending emails, scheduling meetings, revising knowledge articles, or submitting pull requests. We evaluate these tasks using \textbf{state-based verification}: after execution, the resulting system state is automatically compared against the structured ground-truth specification, including required state transitions and critical tool-call parameters. \textcolor{red}{Need to Check!!!!}

\textbf{Workflow subset} covers operational departments, including IT, HR, Customer Service, Shared Services, and Engineering. Tasks in this subset require agents to modify states in external enterprise systems, such as creating incidents, updating HR cases, sending emails, scheduling meetings, revising knowledge articles, or submitting pull requests. Evaluation is performed by extracting the trace events of each agent and comparing them with the structured ground-truth specification. To further verify that actions are truly executed rather than merely proposed, we also collect initial and final snapshots of the relevant databases and examine their diffs for the expected state transitions.

\textbf{Approval subset} covers the Approval Center, whose three specialists (finance, legal, procurement) emit policy-grounded decisions rather than mutate external system state. Given a submitted request, each involved specialist must determine whether the request complies with internal policies and external regulations, cite the rules that justify the decision, and flag cases where the request is missing information required by an applicable rule. We evaluate this subset by \emph{policy-based verification}: each case is scored against a deterministic per-specialist reference decision derived from a curated policy schema, comparing the decision label, the supporting rule citations, and any required-information flags against a reference adjudication.

Both subsets share the same cross-agent delegation protocol, they differ only in the verification target: operational workflow tasks are evaluated by final system state, while approval tasks are evaluated by policy-grounded adjudication.

\subsection{Data Construction}
The two subsets share a common goal: producing executable, verifiable enterprise tasks at scale, but their construction pipelines reflect their different verification semantics. The Workflow subset is built bottom-up from a fixed enterprise tool catalog and seed system state; the Approval subset is built top-down from a structured policy schema of policy rules extracted from authoritative sources.

\subsubsection{Source Materials}
\label{sec:bench:sources}
% \paragraph{Workflow source materials.}
% \textcolor{red}{TODO: NEED to cite EnterpriseOps-Gym and clarify the difference} 
The Workflow subset is built from a \emph{tool catalog} and a \emph{seed database}. The tool catalog defines the enterprise services available to agents, including the permitted tools and parameter schemas for each service. The seed database populates the simulated enterprise systems with concrete business objects, ensuring that all task parameters and ground-truth trajectories refer to valid objects in the environment. The design of the tool catalog and seed database is informed by EnterpriseOps-Gym~\citep{malay2026enterpriseopsgymenvironmentsevaluationsstateful}.

% \paragraph{Approval source materials.}
The Approval subset is built from a curated policy corpus consisting of 60 pages from the GitLab Handbook~\citep{gitlabhandbook} and eleven GDPR articles~\citep{gdpr2016}. The GDPR articles supplement the Handbook with broader data-protection rules. After removing navigation and footer boilerplate, the corpus contains approximately 840K characters of policy text and is processed using the same downstream pipeline.

\subsubsection{Workflow Task Construction}

Workflow tasks are generated category-first. We predefine business categories from common enterprise events and instantiate cases from seed-database parameters, varying the business object, triggering event, agents, and closure artifact. For each case, we generate the ground-truth trajectory from one of 20 manually specified domain templates, each covering one business domain and iterated over five trigger types and five governance rules. The trajectory starts from the agent whose backend owns the trigger and records each step's executing agent, MCP server, tool name, and arguments. Cross-department actions are modeled as explicit delegation events that transfer a subtask and its supporting context to the appropriate downstream role. These events are implemented as role-specific delegation tools, such as \texttt{ask\_<peer>\_by\_http}. A worked-out template appears in Appendix~\ref{app:workflow_template}.

We further verify that each accepted trajectory uses cataloged tools, respects tool ownership, delegates cross-department transitions explicitly, and recovers all argument values from the user instruction. We further verify that each accepted trajectory uses cataloged tools, respects tool ownership, delegates cross-department transitions explicitly, and recovers all argument values from the user instruction.

Cases with a natural multi-stage structure are further partitioned into multi-step tasks. Such cases typically consist of a record-establishment phase, a technical-resolution phase, and a coordination-and-closure phase. Each phase is converted into a sub-task with its own starting agent, its own instruction conditioned on context produced by upstream phases, and its own ground-truth fragment containing only the steps executed within that phase. We require each multi-step task to include at least three cross-agent delegations across the full chain, ensuring that the task cannot be reduced to sequential tool calls by a single agent.

\subsubsection{Approval Task Construction}
\label{sec:bench:approval}
Approval tasks are constructed from structured policy rules. Each task is generated by sampling target rules from a policy schema, constructing a case that triggers those rules, and rendering the case into a submission package with deterministic ground truth.

\textbf{Policy schema construction.}
We transform the policy corpus into a structured schema through four stages: a heading-aware chunker; an LLM classifier (GPT-5.4-Mini) that labels each chunk with its primary role (finance, legal, or procurement) and flags chunks containing approval-relevant policy rules; an LLM extractor (GPT-5.4) that emits rules in a strict JSON schema; and a finalization stage that filters low-quality extractions, deduplicates entries, and normalizes field names.Each extracted rule specifies a primary role, conjunctive typed-field conditions, a decision label in \{\texttt{approve}, \texttt{reject}, \texttt{require\_docs}, \texttt{require\_preapproval}, \texttt{not\_applicable}\}, an approver chain, fulfillment-evidence slugs, and a source citation. The citation includes verbatim policy text and must be a contiguous substring of the source chunk, providing a falsifiable grounding signal that prevents hallucinated rules from entering the schema. The finalized policy schema contains 290 rules: 42 finance rules, 154 legal rules, and 94 procurement rules. A policy schema generation example appears in Appendix~\ref{app:policy_example}.

\textbf{Approval Task synthesis.}
Each Approval task is generated deterministically from the structured policy schema. We first sample one or more target rules and instantiate a case whose fields satisfy their conditions. We then add plausible but non-applicable distractor rules to reduce reliance on shallow pattern matching while preserving the ground truth.
Three optional perturbations operate at the decision level: (i) deleting one of a target rule's predicate fields forces the agent to recognize missing information; (ii) including or omitting a rule's required evidence controls whether its preapproval requirement is discharged; and (iii) choosing target rules across multiple roles forces cross-departmental adjudication.
The constructed case is then rendered into an intake form summarizing the business context and submitted parameters, optional supporting-evidence documents, and a role-specific directive that asks each specialist to review the case against named policy documents. Ground truth is computed by a deterministic \emph{decision engine} that produces per-specialist decisions with supporting rule citations. A task generation example appears in Appendix~\ref{app:task_example}.

The resulting case is rendered into an intake form, optional evidence documents, and role-specific review instructions. A deterministic decision engine then computes the expected per-specialist decision and its supporting rule citations. A task generation example appears in Appendix~\ref{app:task_example}.

Multi-step tasks in the Approval subset repeat this construction across stages that share a single case identifier. Later stages receive a summary of the upstream specialists' decisions in their user prompt, testing whether the agent can maintain stage independence rather than simply carrying upstream outcomes forward.

\section{Evaluation}

\subsection{Multi-Agent System Formulation}

We formulate enterprise multi-agent collaboration as a constrained distributed task execution problem, defined by a four-tuple $(\mathcal{A}, \mathcal{T}, \pi, \mathcal{D})$. The \emph{agent set} $\mathcal{A} = \{a_1, \dots, a_N\}$ ($N{=}11$) is partitioned into operational agents $\mathcal{A}_{\text{op}}$ (8 agents) and approval agents $\mathcal{A}_{\text{appr}}$ (3 agents). The \emph{tool set} $\mathcal{T} = \mathcal{T}_{\text{svc}} \cup \mathcal{T}_{\text{del}}$ consists of enterprise service tools $\mathcal{T}_{\text{svc}}$ distributed across 8 service systems and delegation tools $\mathcal{T}_{\text{del}} = \{t_{\text{del}}^{a_j} : a_j \in \mathcal{A}\}$, where $t_{\text{del}}^{a_j}$ represents the tool for delegating a subtask to agent~$a_j$. The \emph{permission mapping} $\pi\colon \mathcal{A} \rightarrow 2^{\mathcal{T}}$ assigns each agent~$a_i$ its accessible tool set $\pi(a_i) \subseteq \mathcal{T}$; we write $\pi_{\text{svc}}(a_i) = \pi(a_i) \cap \mathcal{T}_{\text{svc}}$ for the enterprise service tool subset. The \emph{delegation mapping} $\mathcal{D}\colon \mathcal{A} \rightarrow 2^{\mathcal{A}}$ specifies the agents to which $a_i$ may delegate. The four-tuple satisfies $\{t_{\text{del}}^{a_j} : a_j \in \mathcal{D}(a_i)\} \subseteq \pi(a_i)$, i.e., each agent's permission set includes the delegation tools for its delegation targets.

Given a natural language instruction~$q$ and a starting agent~$a_{\text{start}}$, the system produces an execution trajectory $\tau = [e_1, \dots, e_K]$, where each step $e_k = (a^{(k)}, t^{(k)}, \theta^{(k)})$ denotes agent~$a^{(k)}$ invoking tool $t^{(k)} \in \pi(a^{(k)})$ with arguments~$\theta^{(k)}$. A step is an enterprise service tool call when $t^{(k)} \in \mathcal{T}_{\text{svc}}$, and a delegation action when $t^{(k)} \in \mathcal{T}_{\text{del}}$. A task~$q$ may comprise multiple subtasks $\mathcal{U}(q) = \{u_1, \dots, u_L\}$, each corresponding to a trajectory segment. We define $\tau|_{a_i, u_l}$ as the subsequence of~$\tau$ consisting of steps belonging to~$u_l$ and executed by~$a_i$, preserving the original order. Evaluation compares~$\tau$ against a reference trajectory~$\tau^*$, supplemented by system state verification.

\subsection{Execution Environment}

\textbf{Enterprise service layer.}\quad The environment comprises 8 enterprise service systems exposing standardized tool interfaces via the MCP protocol, collectively constituting~$\mathcal{T}_{\text{svc}}$. ITSM, HR, CSM, and Gitea serve as core business systems for IT operations, human resources, customer service, and code management respectively; Email, Calendar, Teams, and Drive serve as collaboration systems providing cross-departmental communication and document management capabilities.

% \textbf{Agent layer.}\quad The 11 agents in~$\mathcal{A}$ are organized into six functional departments (Table~\ref{tab:org}). Each agent is powered by an LLM as its reasoning core and equipped with a role-specific system prompt (see Appendix \ref{subsec:inf_prompt}). Agent~$a_i$ has access to~$\pi(a_i)$, which includes its authorized enterprise service tools and delegation tools pointing to agents in~$\mathcal{D}(a_i)$. When $a_i$ invokes~$t_{\text{del}}^{a_j}$, the subtask and its context are sent to~$a_j$, which processes the request independently and returns the result. Delegation may occur recursively, but the chain depth is bounded by a preset limit~$d_{\max}$ (see Appendix \ref{subsec:hyperparam}).

\textbf{Agent layer.}\quad The 11 agents in~$\mathcal{A}$ are organized into six functional departments (Appendix Table~\ref{tab:org}). Each agent uses an LLM as its reasoning core and equipped with a role-specific system prompt (see Appendix \ref{subsec:inf_prompt}). Agent~$a_i$ has access to~$\pi(a_i)$, which includes its authorized enterprise service tools and delegation tools pointing to agents in~$\mathcal{D}(a_i)$. When $a_i$ invokes~$t_{\text{del}}^{a_j}$, the subtask and its context are sent to~$a_j$, which processes the request independently and returns the result. Delegation may occur recursively, but the chain depth is bounded by a preset limit~$d_{\max}$ (see Appendix \ref{subsec:hyperparam}). Notably, each agent operates as an independent service, engaging in peer-to-peer communication and maintaining an isolated memory throughout the entire task execution lifecycle.

\textbf{Instantiation of $\pi_{\text{svc}}$.}\quad $\pi_{\text{svc}}$ is realized through two layers. The first layer is \emph{tool visibility}: core business system tools are assigned to agents by functional domain (e.g., $\pi_{\text{svc}}(\texttt{hr\_service\_specialist})$ includes HR tools), collaboration system tools are available to all $a_i \in \mathcal{A}_{\text{op}}$, and approval agents $a_i \in \mathcal{A}_{\text{appr}}$ have $\pi_{\text{svc}}(a_i)$ restricted to local workspace document reading tools only. The second layer is \emph{identity authentication}: each agent holds an independent identity token for each service system, and calls without a valid token are rejected at the server side, ensuring that even accidentally constructed calls to tools $t \notin \pi_{\text{svc}}(a_i)$ cannot succeed.

\textbf{Data isolation.}\quad Before each task execution, an independent database instance is created for each involved service system and seeded with predefined data, establishing a deterministic initial state~$S_0$. Database instances are fully isolated across tasks, ensuring inter-task independence.

\subsection{Evaluation Procedure}

Each task undergoes four phases. \textbf{(1)~Initialization:} seed the databases and capture the initial state snapshot~$S_0$. \textbf{(2)~Execution:} send instruction~$q$ to~$a_{\text{start}}$, which autonomously invokes tools in $\pi_{\text{svc}}(a_{\text{start}})$ or delegates via $t_{\text{del}}^{a_j}$ to $a_j \in \mathcal{D}(a_{\text{start}})$; downstream agents continue likewise, forming a multi-hop collaboration chain that produces~$\tau$. For multi-step tasks, subtasks are executed sequentially; if a preceding subtask fails, all subsequent subtasks are judged as failed. 
\textbf{(3)~Evidence collection:} capture the final state snapshot~$S_{u_l}^f$ after each subtask~$u_l$ and compute the normalized state difference $\Delta(S_{u_l}^0, S_{u_l}^f)$ by comparing record-level changes table by table, filtering non-semantic fields to retain only business state changes; simultaneously collect execution events from all participating agents and merge them into~$\tau$. For approval tasks, the external service-state difference is empty by design; the emitted approval decision is recorded as the terminal execution event and compared against the reference outcome. \textbf{(4)~Cleanup:} destroy the isolated database instances to ensure the next task starts from a clean state.

\subsection{Judgment Mechanism}

\textbf{Per-agent judgment.}\quad Let $\mathcal{A}(u_l) = \{a_i : \tau^*|_{a_i, u_l} \neq \emptyset\}$ be the set of agents involved in the reference trajectory of subtask~$u_l$. For each $a_i \in \mathcal{A}(u_l)$, the judgment module constructs a three-part input: (1)~$\tau^*|_{a_i, u_l}$, the reference steps that $a_i$ should execute within~$u_l$; (2)~$\tau|_{a_i, u_l}$, the actual execution events of~$a_i$ within~$u_l$; and (3)~$\Delta(S_{u_l}^0, S_{u_l}^f)$, the state difference as objective evidence of whether tool calls produced expected side effects. The judge model compares $\tau^*|_{a_i, u_l}$ against $\tau|_{a_i, u_l}$, tolerating equivalent implementations and reasonable ordering differences, but ruling $\text{pass}(a_i, u_l) = \text{false}$ when key actions are missing or tool arguments contradict the state evidence. Approval agents $a_i \in \mathcal{A}_{\text{appr}}$ are judged by comparing the terminal decision event with the expected approval outcome, including the decision label, supporting rule citations, and missing-information flags.

\textbf{Judgment consistency.}\quad To mitigate single-model judgment bias, the system uses a three-model majority vote: Gemini-3.1-Pro, GPT-5.4, and Claude-Sonnet-4.6 independently determine $\text{pass}(a_i, u_l)$, with the final result decided by majority vote. Appendix~\ref{yizhi} reports the consistency between the voting results and human annotations.

\textbf{Aggregation and metrics.}\quad We define subtask-level pass as $\text{pass}(u_l) = \bigwedge_{a \in \mathcal{A}(u_l)} \text{pass}(a, u_l)$, i.e., subtask~$u_l$ passes iff all involved agents pass. Let $\mathcal{A}_{\text{eval}} = \bigcup_{u_l \in \mathcal{S}} \{(a, u_l) : a \in \mathcal{A}(u_l)\}$ be all evaluated (agent, subtask) pairs, $\mathcal{S} = \bigcup_{q \in \mathcal{Q}} \mathcal{U}(q)$ be all subtasks, and $\mathcal{Q}$ be all tasks. We report three levels of pass rates: $R_{\text{agent}} = |\{(a, u_l) \in \mathcal{A}_{\text{eval}} : \text{pass}(a, u_l)\}| / |\mathcal{A}_{\text{eval}}|$ measures single-step execution accuracy, $R_{\text{subtask}} = |\{u_l \in \mathcal{S} : \text{pass}(u_l)\}| / |\mathcal{S}|$ measures local collaboration success, and $R_{\text{task}} = |\{q \in \mathcal{Q} : \forall\, u_l \in \mathcal{U}(q),\; \text{pass}(u_l)\}| / |\mathcal{Q}|$ measures end-to-end workflow completion.

\section{Experiments}

\subsection{Experimental Setup}

% 我们评估了闭源模型和开源模型，包括 Claude-Sonnet-4.6、Gemini-3.1-Pro-Preview、Gemini-3.1-Flash-Lite-Preview、GPT-5.4、GPT-5-mini、DeepSeek-V4-Pro、DeepSeek-V4-Flash、Qwen3.5-122B-A10B、Qwen3.5-35B-A3B、Qwen3.5-9B、MiniMax-M2.7 和 MiMo-V2-Flash。我们进行了三次运行，并报告平均总体准确率，以及按检索目标模态和音频内容类型划分的准确率。实现细节见附录~\ref{canshu}。

We evaluated both closed-source and open-source models, including Claude-Sonnet-4.6~\citep{anthropic2026sonnet46}, Gemini-3.1-Pro-Preview~\citep{gemini3pro2025}, Gemini-3.1-Flash-Lite-Preview~\citep{gemini3pro2025}, GPT-5.4~\citep{singh2025openai}, GPT-5-mini\citep{singh2025openai}, DeepSeek-V4-Pro~\citep{deepseekai2026deepseekv4}, DeepSeek-V4-Flash~\citep{deepseekai2026deepseekv4}, Qwen3.5-122B-A10B~\citep{qwen35blog}, Qwen3.5-35B-A3B~\citep{qwen35blog}, Qwen3.5-9B~\citep{qwen35blog}, MiniMax-M2.7~\citep{minimax2026m27}, and MiMo-V2-Flash~\citep{coreteam2026mimov2flashtechnicalreport}. We report the overall accuracy, as well as the step-wise and multi-step accuracy for workflow and approval. In addition, we recorded the accuracy of each agent and computed both the average token cost per task and the average token cost for successfully completed tasks. Details of the settings are provided in Appendix \ref{subsec:hyperparam}.

\subsection{Main Results}

\textbf{EntCollabBench remains challenging even for the strongest models.}
As shown in Table~\ref{tab:model-results}, the best overall result is achieved by DeepSeek-V4-Pro with 62.00\% average accuracy, followed by DeepSeek-V4-Flash at 57.33\% and Claude-Sonnet-4.6 at 52.67\%. Most evaluated models remain below 50\%, indicating that realistic enterprise multi-agent collaboration is far from solved.

\textbf{End-to-end collaboration is substantially harder than solving individual subtasks.}
As shown in Table~\ref{tab:model-results}, on multi-step workflow tasks, DeepSeek-V4-Pro reaches 78.33\% subtask accuracy but only 50.00\% task accuracy, while Claude-Sonnet-4.6 drops from 69.17\% to 50.00\%. This gap suggests that errors accumulate across delegation chains, where a single routing, execution, or communication failure can cause the full workflow to fail.

\textbf{Approval tasks are easier in isolation but still difficult in multi-step settings.}
As shown in Table~\ref{tab:model-results}, in single-step approval tasks, DeepSeek-V4-Flash reaches 80.00\%, and both Claude-Sonnet-4.6 and DeepSeek-V4-Pro reach 78.75\%. However, the best multi-step approval task accuracy is only 40.00\%, showing that policy reasoning must be combined with evidence preservation, role-specific judgment, and consistent cross-stage coordination.

\textbf{Role-level success is much higher than full-task success.}
Tables~\ref{tab:role-performance-single} and~\ref{tab:role-performance-multi} show that strong models often exceed 80\% average accuracy at the role level, while their end-to-end task accuracy is much lower. This indicates that the main bottleneck is not isolated tool execution, but cross-agent routing, context transmission, and coordination under permission isolation.

\textbf{Higher token usage does not necessarily translate into better collaboration.}
Tables \ref{tab:token-stats-success} and \ref{tab:token-stats} show that multi-step tasks consume substantially more tokens, especially in the Workflow Track. However, models with larger token usage are not always more accurate, suggesting that successful enterprise collaboration depends more on concise delegation, faithful parameter preservation, and effective planning than on longer context alone.

% green for mul / retrieval-type columns
\definecolor{mcphigh}{RGB}{51,255,51}
% blue for non-mul / content-type columns
\definecolor{apprhigh}{RGB}{51,158,255}
% red for avg
\definecolor{avghigh}{RGB}{255,90,90}

\newcommand{\colormcp}[1]{\cellcolor{mcphigh!#1}}
\newcommand{\colorappr}[1]{\cellcolor{apprhigh!#1}}
\newcommand{\coloravg}[1]{\cellcolor{avghigh!#1}}

\renewcommand{\arraystretch}{1.2}

\begin{table*}[t]
\centering
\caption{Experimental results of closed-source and open-source models on EntCollabBench. Avg. is the task-level average.}
\label{tab:model-results}
\setlength{\tabcolsep}{0pt}
\renewcommand{\arraystretch}{1.25} 
\resizebox{0.9\textwidth}{!}{
\begin{tabular}{l @{\hspace{12pt}} *{9}{>{\centering\arraybackslash}m{0.110\textwidth}}}
\toprule
\multirow{2}{*}[-1.0ex]{\textbf{Model}} &
\multirow{2}{*}[-1.0ex]{\textbf{workflow}} &
\multicolumn{2}{c}{\makebox[0pt][c]{\textbf{workflow multi-task}}} &
\multirow{2}{*}[-1.0ex]{\textbf{\makecell{workflow\\avg.}}} &
\multirow{2}{*}[-1.0ex]{\textbf{approval}} &
\multicolumn{2}{c}{\makebox[0pt][c]{\textbf{approval multi-task}}} &
\multirow{2}{*}[-1.0ex]{\textbf{\makecell{approval\\avg.}}} &
\multirow{2}{*}[-1.0ex]{\textbf{avg.}} \\
\cmidrule{3-4} \cmidrule{7-8}
& & \textbf{subtask} & \makecell{\textbf{task}} & & & \textbf{subtask} & \makecell{\textbf{task}} & & \\
\midrule
\multicolumn{10}{l}{\textbf{Closed-source Models}} \\
\addlinespace[3pt]
\hdashline
\addlinespace[3pt]
Claude-Sonnet-4.6 & \colormcp{63}42.50 & \colormcp{91}69.17 & \colormcp{71}50.00 & \colormcp{80}44.00 & \colorappr{99}78.75 & \colorappr{73}52.73 & \colorappr{55}35.00 & \colorappr{100}70.00 & \coloravg{86}52.67 \\
Gemini-3.1-Pro-Preview & \colormcp{63}41.88 & \colormcp{85}63.33 & \colormcp{66}45.00 & \colormcp{78}42.50 & \colorappr{84}63.75 & \colorappr{69}49.09 & \colorappr{55}35.00 & \colorappr{85}58.00 & \coloravg{79}47.67 \\
Gemini-3.1-Flash-Lite-Preview & \colormcp{22}1.88 & \colormcp{50}29.17 & \colormcp{23}2.50 & \colormcp{23}2.00 & \colorappr{72}52.50 & \colorappr{62}41.82 & \colorappr{40}20.00 & \colorappr{70}46.00 & \coloravg{33}16.67 \\
GPT-5.4 & \colormcp{61}40.62 & \colormcp{63}41.67 & \colormcp{25}5.00 & \colormcp{65}33.50 & \colorappr{88}67.50 & \colorappr{65}45.45 & \colorappr{50}30.00 & \colorappr{87}60.00 & \coloravg{71}42.33 \\
GPT-5-mini & \colormcp{40}19.38 & \colormcp{64}43.33 & \colormcp{28}7.50 & \colormcp{43}17.00 & \colorappr{88}67.50 & \colorappr{78}58.18 & \colorappr{60}40.00 & \colorappr{90}62.00 & \coloravg{56}32.00 \\
\midrule
\multicolumn{10}{l}{\textbf{Open-source Models}} \\
\addlinespace[3pt]
\hdashline
\addlinespace[3pt]
DeepSeek-V4-Pro & \colormcp{83}61.25 & \colormcp{100}78.33 & \colormcp{71}50.00 & \colormcp{100}59.00 & \colorappr{99}78.75 & \colorappr{62}41.82 & \colorappr{45}25.00 & \colorappr{97}68.00 & \coloravg{100}62.00 \\
DeepSeek-V4-Flash & \colormcp{74}52.50 & \colormcp{91}70.00 & \colormcp{66}45.00 & \colormcp{89}51.00 & \colorappr{100}80.00 & \colorappr{67}47.27 & \colorappr{50}30.00 & \colorappr{100}70.00 & \coloravg{93}57.33 \\
Qwen3.5-122B-A10B & \colormcp{51}30.63 & \colormcp{78}56.67 & \colormcp{35}15.00 & \colormcp{57}27.50 & \colorappr{71}51.25 & \colorappr{51}30.91 & \colorappr{40}20.00 & \colorappr{68}45.00 & \coloravg{58}33.33 \\
Qwen3.5-35B-A3B & \colormcp{46}25.62 & \colormcp{74}52.50 & \colormcp{43}22.50 & \colormcp{54}25.00 & \colorappr{72}52.50 & \colorappr{45}25.45 & \colorappr{25}5.00 & \colorappr{66}43.00 & \coloravg{54}31.00 \\
Qwen3.5-9B & \colormcp{20}0.00 & \colormcp{20}0.00 & \colormcp{20}0.00 & \colormcp{20}0.00 & \colorappr{48}27.50 & \colorappr{36}16.36 & \colorappr{25}5.00 & \colorappr{40}23.00 & \coloravg{20}7.67 \\
MiniMax-M2.7 & \colormcp{40}19.38 & \colormcp{75}54.17 & \colormcp{35}15.00 & \colormcp{45}18.50 & \colorappr{65}45.00 & \colorappr{41}20.75 & \colorappr{25}5.00 & \colorappr{58}37.00 & \coloravg{45}24.67 \\
MiMo-V2-Flash & \colormcp{44}23.75 & \colormcp{63}41.67 & \colormcp{33}12.50 & \colormcp{49}21.50 & \colorappr{29}8.75 & \colorappr{20}0.00 & \colorappr{20}0.00 & \colorappr{20}7.00 & \coloravg{33}16.67 \\
\bottomrule
\end{tabular}
}
\vspace{-13pt}
\end{table*}

\subsection{Further Analysis}

To complement aggregate metrics, we analyze representative execution traces to uncover recurring mechanisms behind successful and failed runs. We organize the analysis around enterprise-collaboration behaviors and note model-specific manifestations where especially pronounced.

\textbf{Role difficulty is strongly affected by position in the delegation chain.}
The Knowledge Base Specialist is the weakest operational role, but much of this weakness comes from its frequent position at the end of delegation chains. When it is the starting agent, its pass rate is considerably higher; when it is downstream, failures often originate from missing delegation, incomplete content, or incorrect upstream context. By contrast, Developer and QA roles perform better because repository operations are more structured and often occur earlier in the workflow. Details are provided in Appendix~\ref{app:case_role_position}.

\textbf{Multi-step tasks fail through prefix decay and final handoff errors.}
Multi-step tasks show clear degradation as the chain proceeds. Approval workflows mainly drop at the second step, while workflow tasks often fail at the final step. Details are provided in Appendix~\ref{app:case_prefix_decay}. This pattern is especially visible in Qwen3.5-122B-A10B, which often completes earlier subtasks but fails when the last step requires binding previous artifacts, the current role, and the downstream role.

% (See Appendix~\ref{app:qwen_error}).

\textbf{Collaboration tools become bottlenecks during workflow closure.}
Email, Calendar, and Teams operations often cause failures, despite appearing simpler than core business systems. Agents may misuse sender identities, confuse account keys with email addresses, misformat payloads, or resolve pronouns as literal team/channel names. GPT-5.4 shows this issue most prominently in the \texttt{collaboration\_ops\_specialist} role: many multi-step workflows fail at the final communication stage, contributing to its relatively low accuracy on workflow multi-task. See Appendix~\ref{app:case_collaboration_tools}.

\textbf{Delegation failures remain a central source of end-to-end errors.}
Agents frequently omit required delegation, pass insufficient context, or delegate before prerequisites are ready. Downstream agents may receive tasks without the needed article body, branch, file, or business object, making failure unavoidable even when the downstream role itself is capable. These failures explain why role-level accuracy is much higher than full-task accuracy. Details are provided in Appendix~\ref{app:case_delegation}.

\textbf{Stateful database operations trigger incorrect fallback actions.}
A recurring Workflow Track failure is that agents perform a semantically similar but incorrect database operation after failing to locate the target record. For example, when asked to update an existing incident or knowledge article, agents may search with the wrong identifier field, fail to retrieve the record, and then create a new record instead. This behavior appears across multiple models and is especially harmful because it leaves persistent but wrong enterprise state. Details are provided in Appendix~\ref{app:case_database_fallback}.

\textbf{Tool calls often fail at the parameter-semantics level.}
Many failures occur after the correct tool family is selected. Models choose wrong enum values, mix up relationship labels such as \texttt{applied} and \texttt{suggested}, assign work through incorrect user fields, or set the wrong task status/type. These errors show that enterprise tool use requires parameter-level grounding beyond high-level action selection. Details are provided in Appendix~\ref{app:case_parameter_semantics}.

\textbf{Higher reliability can require much higher coordination cost.}
DeepSeek-V4-Pro achieves strong accuracy, but its successful runs often involve many more trace events and substantially higher token usage. The model appears to execute conservatively, repeatedly checking state and coordinating with downstream agents before finalizing actions. This improves robustness but exposes a cost-efficiency trade-off for enterprise collaboration. Details are provided in Appendix~\ref{app:case_coordination_cost}.

\textbf{Approval workflows expose weak decision commitment.}
In approval tasks, some models retrieve relevant policy evidence but fail to convert it into a final decision. MiMo-V2-Flash shows the most severe form of this behavior, repeatedly reading the same policy documents until token usage grows sharply or the context window is exhausted. This suggests that approval agents need not only retrieval ability, but also stopping criteria and decision discipline. See Appendix~\ref{app:case_approval_loops} for details.

\textbf{Small models are much weaker on executable workflow tasks.}
Small models perform substantially worse on MCP workflow tasks than on approval tasks. Their main bottleneck is executable tool grounding: some stop after listing tools or reading schemas, while others choose plausible tools but fail to produce valid JSON arguments. This suggests that state-changing workflows impose stricter interface and parameter requirements than policy-review tasks. Details are provided in Appendix~\ref{app:case_small_models_mcp}.

\textbf{Some failures occur before real tool execution.}
MiniMax-M2.7 occasionally outputs textual pseudo-tool calls rather than valid executable calls. The model appears to intend an action, but no real tool invocation is recorded in the trace. This reflects weak grounding between natural-language planning and executable tool-use format. See Appendix~\ref{app:case_pseudo_actions} for details.

% Overall, these case studies show that EntCollabBench exposes both shared collaboration bottlenecks and model-specific manifestations of those bottlenecks.

% \section{Conclusion}

% We presented \textsc{EntCollabBench}, a benchmark for evaluating enterprise multi-agent collaboration under role specialization, permission isolation, and cross-departmental delegation. It includes workflow tasks that require state changes in enterprise systems and approval tasks that require policy-grounded decisions. Experiments show that current LLM agents still struggle with end-to-end organizational workflows. While agents often succeed on local role-specific actions, performance drops when tasks require routing, delegation, context transfer, and final-stage coordination. These results highlight the need to evaluate enterprise agents beyond single-agent tool use. \textsc{EntCollabBench} provides a reproducible testbed for measuring and improving agents that aim to operate in realistic organizational environments.

\section{Conclusion}

We presented \textsc{EntCollabBench}, a benchmark for evaluating enterprise multi-agent collaboration under role specialization, permission isolation, and cross-departmental delegation. Experiments show that current LLM agents often handle local role-specific actions, but struggle with end-to-end workflows requiring routing, context transfer, and final-stage coordination. \textsc{EntCollabBench} provides a reproducible testbed for measuring and improving agents in realistic organizational environments.

% \newpage

% \section*{Contributions}

% \textbf{Authors} 
% \quad
% Yuzheng Cai\textsuperscript{\rm 1,2*} \quad Siqi Cai\textsuperscript{\rm 1*}\quad Yuchen Shi\textsuperscript{\rm 1*}\quad Zihan Xu\textsuperscript{\rm 1*} \quad Lichao Chen\textsuperscript{\rm 1,3} \quad Yulei Qin\textsuperscript{\rm 1} \quad Xiaoyu Tan\textsuperscript{\rm 1}\quad Gang Li\textsuperscript{\rm 1} \quad Zongyi Li\textsuperscript{\rm 1} \quad Haojia Lin\textsuperscript{\rm 1} \quad Yong Mao\textsuperscript{\rm 1} \quad Ke Li\textsuperscript{\rm 1\faEnvelopeO} \quad Xing Sun\textsuperscript{\rm 1}

% \textbf{Affiliations} 
% \quad
% \textsuperscript{\rm 1}Tencent Youtu Lab\quad \textsuperscript{\rm 2}Fudan University\quad \textsuperscript{\rm 3}Xiamen University

% \textbf{*Equal Contributions} 
% \quad Yuzheng Cai \quad Siqi Cai \quad Yuchen Shi \quad Zihan Xu

\setcitestyle{numbers,square}
\bibliography{citation}

\newpage
%%%%%%%%%%%%%%%%%%%%%%%%%%%%%%%%%%%%%%%%%%%%%%%%%%%%%%%%%%%%

\newpage
\EnableTOC
\clearpage
\appendix

\section*{Appendix}
\begingroup
\setcounter{tocdepth}{2}  % 0=只显示section, 1=到subsection, 2=到subsubsection
\tableofcontents
\endgroup

\section{Agent Roster}
\label{app:org}

Table~\ref{tab:org} lists all eleven agents in the EntCollabBench organization, grouped by department. Each row reports the agent identifier used in task specifications and ground-truth trajectories, the persona name surfaced to the agent's own system prompt, and the agent's dedicated service scope. The eight operational agents additionally share four common collaboration services (Teams, Email, Calendar, Drive) which are not repeated per row. Cross-agent handoffs are issued through the typed delegation primitive \texttt{ask\_<agent>\_by\_http}, where \texttt{<agent>} is the target identifier in this table.

\begin{table}[h]
\centering\small
\setlength{\tabcolsep}{4pt}
\begin{tabular}{llll}
\toprule
\textbf{Department} & \textbf{Agent identifier} & \textbf{Persona} & \textbf{Dedicated services} \\
\midrule
\multirow{2}{*}{IT}
  & \texttt{it\_service\_desk\_l1}        & Ivan Park    & ITSM \\
  & \texttt{it\_change\_engineer}         & Nina Patel   & ITSM \\
\midrule
Human Resources    & \texttt{hr\_service\_specialist}       & Helen Zhou    & HR \\
\midrule
Customer Service   & \texttt{customer\_support\_specialist} & Carlos Mendez & CSM \\
\midrule
\multirow{2}{*}{Shared Services}
  & \texttt{knowledge\_base\_specialist}   & Priya Nair   & ITSM, HR, CSM (kb tools) \\
  & \texttt{collaboration\_ops\_specialist} & Olivia Chen  & --- (common tools only) \\
\midrule
\multirow{2}{*}{Engineering}
  & \texttt{developer\_engineer}          & Ethan Walker & Gitea \\
  & \texttt{qa\_test\_engineer}           & Mia Kim      & Gitea \\
\midrule
\multirow{3}{*}{Approval Center}
  & \texttt{finance\_approval\_specialist}     & Sophia Lin & local workspace docs \\
  & \texttt{legal\_approval\_specialist}       & Daniel Wu  & local workspace docs \\
  & \texttt{procurement\_approval\_specialist} & Grace Liu  & local workspace docs \\
\bottomrule
\end{tabular}
\caption{Full EntCollabBench agent roster.}
\label{tab:org}
\end{table}

\section{Policy Schema Example}
\label{app:policy_example}

To make the policy schema of Section~\ref{sec:bench:approval} concrete, we show one finalized rule end to end. The rule below is extracted from the GitLab Handbook by the four-stage pipeline (chunker $\rightarrow$ classifier $\rightarrow$ extractor $\rightarrow$ finalize) and is one of the two target rules used by the task example in Appendix~\ref{app:task_example}.

\begin{tcolorbox}[
    title=\textbf{Example: Rule \texttt{LEG-EVAL-0001} (\textit{Evaluation Agreement Approval})},
    fonttitle=\bfseries,
    breakable,
    fontupper=\small
]

\textbf{Identity.}
\begin{itemize}
\item \texttt{rule\_id}: \texttt{LEG-EVAL-0001}
\item \texttt{primary\_role}: \texttt{legal}
\item \texttt{category}: \texttt{evaluation\_agreement\_approval}
\item \texttt{severity}: \texttt{mandatory}
\end{itemize}

\textbf{Conditions} (conjunctive predicates over typed fields; all must hold for the rule to fire):
\begin{itemize}
\item \texttt{customer\_declines\_trial\_process} \texttt{==} \texttt{true}
\item \texttt{requested\_agreement\_type} \texttt{==} \texttt{"Evaluation Agreement"}
\end{itemize}

\textbf{Decision and approver chain.}
\begin{itemize}
\item \texttt{decision}: \texttt{require\_preapproval}
\item \texttt{approver\_chain}: [\texttt{"Area Sales Manager or higher"}]
\item \texttt{fulfillment\_evidence}: [\texttt{area\_sales\_manager\_or\_higher\_preapproval}]
\end{itemize}
The decision is \texttt{require\_preapproval} by default; the rule engine promotes it to \texttt{approve} when an evidence document keyed by the listed slug is present in the case submission.

\textbf{Cross-domain links} (added by the finalize stage):
\begin{itemize}
\item \texttt{cross\_refs}: [\texttt{PROC-AGREE-0001}]
\end{itemize}
The link records that this legal rule \emph{references} a procurement rule on related agreement handling, computed by an LLM pass over rule pairs that share categories or fields.

\textbf{Source citation} (verbatim substring of the original chunk; non-substring extractions are dropped during finalize):
\begin{itemize}
\item \texttt{source.doc\_id}: \texttt{gitlab/legal\_customer\_negotiations}
\item \texttt{source.section}: ``Sales Guide $|$ Collaborating with GitLab Legal $>$ OPERATIONAL $>$ Request for a Trial or Evaluation Agreement''
\item \texttt{source.verbatim}:\\
\textit{``If a customer declines the trial process and is adamant to have a separate Evaluation Agreement, the sales team member or solutions architect should: Open a Legal Request to request an Evaluation Agreement with Request Form. The Legal Request should (i) include a request for approval from the Area Sales Manager or higher; and (ii) set forth applicable details to complete the Request Form, such as customer contact information, length of evaluation, number of users, etc.''}
\end{itemize}

\end{tcolorbox}

The schema-level guarantees illustrated by this rule --- typed conjunctive predicates, an enumerated decision class, named approver chain, evidence slugs, finalized cross-domain links, and a verbatim-grounded citation --- are uniform across all 290 rules in the corpus, so any sampled subset can be fed to the deterministic decision engine without rule-specific glue.

\section{Approval Task Example}
\label{app:task_example}

\sloppy

We walk through one approval task end to end to show how target-rule sampling, distractor injection, case construction, submission rendering, and rule-engine ground truth fit together. The example is task \texttt{T-0001} (\texttt{case\_id}: \texttt{CONT-2026-0001}), a single-step cross-domain case that fires one legal rule and one procurement rule simultaneously.

\paragraph{Step 1 --- Target rule sampling.}
The synthesis pipeline samples two target rules drawn from different roles, exercising the cross-departmental adjudication perturbation:

\begin{tcolorbox}[
    title=\textbf{Sampled target rules},
    fonttitle=\bfseries,
    breakable,
    fontupper=\small
]
\begin{itemize}
\item \textbf{\texttt{LEG-EVAL-0001}} --- category \textit{evaluation\_agreement\_approval}, role \texttt{legal}, decision \texttt{require\_preapproval}. Predicates:
  \begin{itemize}
  \item \texttt{customer\_declines\_trial\_process} $=$ \texttt{true}
  \item \texttt{requested\_agreement\_type} $=$ \texttt{"Evaluation Agreement"}
  \end{itemize}
  Fulfilled by \texttt{area\_sales\_manager\_or\_higher\_preapproval}. Full definition in Appendix~\ref{app:policy_example}.
\item \textbf{\texttt{PROC-BGSCRN-0007}} --- category \textit{contractor\_background\_screening}, role \texttt{procurement}, decision \texttt{require\_docs}. Predicates:
  \begin{itemize}
  \item \texttt{worker\_type} $=$ \texttt{"contingent\_worker"}
  \item \texttt{returning\_to\_service\_at\_gitlab} $=$ \texttt{true}
  \item \texttt{days\_since\_contract\_completion} $\leq$ \texttt{90}
  \end{itemize}
  Fulfilled by \texttt{contractor\_background\_screening}.
\end{itemize}
\end{tcolorbox}

\paragraph{Step 2 --- Case construction.}
Field values are reverse-engineered from the predicates of all target rules so that every condition is satisfied; remaining business fields (project name, applicant department, application date) are filled from a fixture pool.

\begin{tcolorbox}[
    title=\textbf{Constructed case parameters (\texttt{case\_id}: \texttt{CONT-2026-0001})},
    fonttitle=\bfseries,
    breakable,
    fontupper=\small
]
\begin{itemize}
\item \texttt{worker\_type} $=$ \texttt{"contingent\_worker"}
\item \texttt{returning\_to\_service\_at\_gitlab} $=$ \texttt{true}
\item \texttt{days\_since\_contract\_completion} $=$ \texttt{90}
\item \texttt{customer\_declines\_trial\_process} $=$ \texttt{true}
\item \texttt{requested\_agreement\_type} $=$ \texttt{"Evaluation Agreement"}
\item \texttt{extension\_beyond\_initial\_term} $=$ \texttt{true}
\end{itemize}
Business header: \texttt{project\_name} $=$ \texttt{"Horizon Marketing Refresh"}; \texttt{applicant\_department} $=$ \texttt{"Infrastructure Operations"}; \texttt{application\_date} $=$ \texttt{"2026-05-03"}.
\end{tcolorbox}

\paragraph{Step 3 --- Distractor injection.}
Distractor rules from the same legal/procurement neighborhood are added to tighten the read-time decision boundary. Their predicates are pinned at near-miss values so they almost fire on the case but ultimately do not, leaving the ground truth unchanged:

\begin{tcolorbox}[
    title=\textbf{Distractor rules (do not fire)},
    fonttitle=\bfseries,
    breakable,
    fontupper=\small
]
\begin{itemize}
\item \textbf{\texttt{PROC-CW-0001}} --- category \textit{contractor\_extension\_approval}.
  \begin{itemize}
  \item Rule requires: \texttt{worker\_type} $=$ \texttt{"staff\_augmentation\_worker"}
  \item Case has: \texttt{worker\_type} $=$ \texttt{"contingent\_worker"}
  \item Near-miss: the rule misses on \texttt{worker\_type}, although the case's \texttt{extension\_beyond\_initial\_term} $=$ \texttt{true} would otherwise align with a later predicate.
  \end{itemize}
\item \textbf{\texttt{PROC-CW-0002}} --- variant of the same category with the same near-miss on \texttt{worker\_type}.
\end{itemize}
\end{tcolorbox}

\paragraph{Step 4 --- Submission rendering.}
The synthesizer renders the case into a self-contained submission package: a shared intake form, one evidence document per fulfilled rule, and a role-specific directive that names which specialists must adjudicate which sub-review.

\begin{tcolorbox}[
    title=\textbf{Submission artifacts (excerpts)},
    fonttitle=\bfseries,
    breakable,
    fontupper=\small
]

\textbf{(a) Approval intake form (excerpt).}
\begin{quote}
\textit{``Approval Intake Form for case\_id CONT-2026-0001. Project Name: Horizon Marketing Refresh. Applicant Department: Infrastructure Operations. Application Date: 2026-05-03. The submission flags Contractor Background Screening as the operative review area\dots\\
Submitted parameters: worker\_type = contingent\_worker; returning\_to\_service\_at\_gitlab = True; days\_since\_contract\_completion = 90; customer\_declines\_trial\_process = True; requested\_agreement\_type = Evaluation Agreement; extension\_beyond\_initial\_term = True.''}
\end{quote}

\textbf{(b) Pre-approval evidence (\texttt{LEG-EVAL-0001} fulfillment).}
\begin{quote}
\textit{From: Area Sales Manager or higher. To: Infrastructure Operations Intake. Subject: Evaluation Agreement Approval --- pre-approval --- CONT-2026-0001.\\
``As Area Sales Manager or higher within the legal domain for the cited control I pre-approve this routing on the documented parameters.''}
\end{quote}

\textbf{(c) Background-check evidence (\texttt{PROC-BGSCRN-0007} fulfillment).}
\begin{quote}
\textit{Subject: contractor background screening --- completed for the contractor background screening workflow on Horizon Marketing Refresh.\\
``This report satisfies the documentation requirement keyed by contractor\_background\_screening for the routed approval.''}
\end{quote}

\textbf{(d) Directive (delegation prompt to the orchestrator).}
\begin{quote}
\textit{``case\_id CONT-2026-0001. Project Horizon Marketing Refresh is submitted by Infrastructure Operations on 2026-05-03\dots\\
First, ask legal\_approval\_specialist to perform Contractor Background Screening --- Legal \& Regulatory Review; the review should read the intake form, the background-check report, the pre-approval email, and the legal material-review policy document\dots\\
Then, ask procurement\_approval\_specialist to perform Contractor Background Screening --- Procurement \& Vendor Review; the review should read the same evidence and the procurement background-screening policy document\dots''}
\end{quote}

\end{tcolorbox}

\paragraph{Step 5 --- Ground-truth computation.}
The deterministic decision engine evaluates the case against the full policy schema. Two rules fire (one per role); both have their fulfillment evidence present in the submission, so the engine promotes \texttt{require\_preapproval} / \texttt{require\_docs} to \texttt{approve}. The finance specialist has no firing rule and resolves to \texttt{not\_applicable}.

\begin{tcolorbox}[
    title=\textbf{Ground truth (\texttt{ground\_truth\_approval\_results})},
    fonttitle=\bfseries,
    breakable,
    fontupper=\small
]
\begin{itemize}
\item \textbf{\texttt{finance\_approval\_specialist}}
  \begin{itemize}
  \item Decision: \texttt{not\_applicable}
  \item Citations: \texttt{[]}
  \item Reason: no finance rule is implicated by the case parameters.
  \end{itemize}
\item \textbf{\texttt{legal\_approval\_specialist}}
  \begin{itemize}
  \item Decision: \texttt{approve}
  \item Citations: \texttt{[LEG-EVAL-0001]}
  \item Reason: the rule fires on the case parameters; the pre-approval evidence keyed by \texttt{area\_sales\_manager\_or\_higher\_preapproval} discharges the pre-approval requirement, promoting the rule from \texttt{require\_preapproval} to \texttt{approve}.
  \end{itemize}
\item \textbf{\texttt{procurement\_approval\_specialist}}
  \begin{itemize}
  \item Decision: \texttt{approve}
  \item Citations: \texttt{[PROC-BGSCRN-0007]}
  \item Reason: the rule fires on the case parameters; the background-check evidence keyed by \texttt{contractor\_background\_screening} discharges the document requirement, promoting the rule from \texttt{require\_docs} to \texttt{approve}.
  \end{itemize}
\end{itemize}
\end{tcolorbox}

\paragraph{Step 6 --- Final task record.}
The five steps above are serialized into a single structured record that is appended to the released \texttt{tasks.json} file. The abridged record below shows the fields actually consumed by the runtime: the input parameters, the metadata used by analyses (e.g.\ which rules fire and which are distractors), the rendered submission package, and the rule-engine ground truth. Long string fields (e.g.\ \texttt{user\_prompt}, per-specialist \texttt{rationale}) are elided because they appear verbatim in earlier steps.

\begin{tcolorbox}[
    title=\textbf{Released task record (\texttt{tasks.json}, abridged)},
    fonttitle=\bfseries,
    breakable,
    fontupper=\ttfamily\footnotesize
]
\{\\
\hspace*{1em}"task\_id": "T-0001",\\
\hspace*{1em}"type": "contractor\_background\_screening",\\
\hspace*{1em}"description": "Contractor Background Screening review for\\
\hspace*{2.5em}Infrastructure Operations (case\_id CONT-2026-0001).",\\
\hspace*{1em}"task\_list": [\{\\
\hspace*{2em}"sub\_index": 1,\\
\hspace*{2em}"begin\_agent": "approval\_orchestrator",\\
\hspace*{2em}"user\_prompt": "\textrm{\textit{(see Step~4(d) above)}}",\\
\hspace*{2em}"request\_fields": \{\\
\hspace*{3em}"worker\_type": "contingent\_worker",\\
\hspace*{3em}"returning\_to\_service\_at\_gitlab": true,\\
\hspace*{3em}"days\_since\_contract\_completion": 90,\\
\hspace*{3em}"customer\_declines\_trial\_process": true,\\
\hspace*{3em}"requested\_agreement\_type": "Evaluation Agreement",\\
\hspace*{3em}"extension\_beyond\_initial\_term": true\\
\hspace*{2em}\},\\
\hspace*{2em}"satisfied\_rule\_ids":  ["LEG-EVAL-0001", "PROC-BGSCRN-0007"],\\
\hspace*{2em}"distractor\_rule\_ids": ["PROC-CW-0001", "PROC-CW-0002"],\\
\hspace*{2em}"involved\_roles":      ["legal", "procurement"],\\
\hspace*{2em}"submission\_package": [\\
\hspace*{3em}\{"evidence\_id": "E-001", "kind": "form"\},\\
\hspace*{3em}\{"evidence\_id": "E-002", "kind": "background\_check"\},\\
\hspace*{3em}\{"evidence\_id": "E-003", "kind": "preapproval\_email"\}\\
\hspace*{2em}],\\
\hspace*{2em}"ground\_truth\_approval\_results": \{\\
\hspace*{3em}"finance\_approval\_specialist":\\
\hspace*{4em}\{"decision": "not\_applicable", "rule\_citations": []\},\\
\hspace*{3em}"legal\_approval\_specialist":\\
\hspace*{4em}\{"decision": "approve", "rule\_citations": ["LEG-EVAL-0001"]\},\\
\hspace*{3em}"procurement\_approval\_specialist":\\
\hspace*{4em}\{"decision": "approve", "rule\_citations": ["PROC-BGSCRN-0007"]\}\\
\hspace*{2em}\}\\
\hspace*{1em}\}]\\
\}
\end{tcolorbox}

\paragraph{What the agent must do.}
The agent must (i) recognize that \texttt{LEG-EVAL-0001} and \texttt{PROC-BGSCRN-0007} are the two firing rules and that \texttt{PROC-CW-0001} / \texttt{PROC-CW-0002} are near-miss distractors that do not apply, (ii) verify that the corresponding fulfillment evidence is present in the submission package, and (iii) emit per-specialist decisions with rule citations matching the engine's ground truth. The finance specialist must correctly emit \texttt{not\_applicable} with empty citations, since no finance rule is implicated.

\section{Workflow Task Template Example}
\label{app:workflow_template}

\sloppy

The Workflow subset is generated from a fixed library of 20 domain templates, one per business domain. Each template is a Python function that takes a render context (trigger key, governance key, dates, timezone, fixture indices) and emits a \texttt{TaskDraft} containing (i) a parameter-faithful natural-language instruction (in English, with a Chinese translation), and (ii) an ordered tool sequence over typed enterprise services that drives the ground-truth trajectory and the verifier's expected state changes. Each template is invoked under the cross-product of 5 triggers and 5 governance rules, yielding up to 500 raw task drafts per generation pass before deduplication and quality filtering.

We illustrate the structure with one template, \texttt{build\_employee\_onboarding\_provisioning}, which renders an HR-led onboarding case that crosses People Ops, IT Service Desk, the hiring department, and finance shared services.

\begin{tcolorbox}[
    title=\textbf{Example template: employee onboarding provisioning},
    fonttitle=\bfseries,
    breakable,
    fontupper=\small
]

\textbf{Trigger fixtures} (one of five sampled by trigger\_index):
\begin{itemize}
\item \textit{monitoring\_alert} --- ``The onboarding readiness dashboard on $\langle$trigger\_date$\rangle$ shows that the $\langle$role$\rangle$ start package for the $\langle$office$\rangle$ is missing identity and equipment tasks.''
\item \textit{employee\_request} --- ``The hiring coordination mailbox asked on $\langle$trigger\_date$\rangle$ to accelerate onboarding for the $\langle$role$\rangle$ joining the $\langle$department$\rangle$ team in the $\langle$office$\rangle$.''
\item \textit{audit\_finding} --- ``An internal onboarding audit opened on $\langle$trigger\_date$\rangle$ found that the approved $\langle$role$\rangle$ requisition for the $\langle$office$\rangle$ has no complete cross-functional provisioning trail.''
\item \textit{contract\_date\_node} --- start-milestone variant keyed to action\_date.
\item \textit{external\_partner\_event} --- payroll-feed handoff variant.
\end{itemize}

\textbf{Governance fixtures} (one of five sampled by governance\_index):
\begin{itemize}
\item \textit{sla\_deadline} --- the day-one provisioning package must close by the action date.
\item \textit{budget\_cost} --- Finance Controller sign-off above \$3{,}200.
\item \textit{approval\_chain} --- People Ops Director and IT Operations Manager.
\item \textit{compliance\_legal} --- new-joiner access segregation.
\item \textit{continuity\_min\_disruption} --- preserve the team's start-of-quarter operations.
\end{itemize}

\textbf{Parameterized roles, departments, and offices} (drawn from per-template fixture pools):
\begin{itemize}
\item Roles: APAC Revenue Operations Analyst; EMEA Support Enablement Specialist; North America Procurement Coordinator; Shared Services Payroll Analyst; Cloud Operations Planner.
\item Departments: Revenue Operations; Customer Support; Procurement; Finance Shared Services; Cloud Operations.
\item Offices: New York / Los Angeles / London / Shanghai / Singapore Hub.
\end{itemize}

\textbf{Tool sequence} (the structured trajectory; each step is parameter-checked against the seed database and is owned by a specific MCP server):
\begin{itemize}
\item \texttt{frappe.get\_job\_openings}
\item \texttt{eops\_hr.list\_hr\_cases}
\item \texttt{eops\_hr.create\_new\_hr\_case} (or \texttt{update\_hr\_case} if the requisition already has an open case)
\item \texttt{eops\_itsm.create\_incident}
\item \texttt{eops\_teams.create\_group}
\item \texttt{eops\_email.send\_email} \textit{or} \texttt{eops\_teams.send\_message} (chosen by context)
\end{itemize}

\textbf{Rendered instruction} (one filled instance, after the parameter-faithful skeleton is rewritten by GPT-5.5 into stakeholder voice):
\begin{quote}
\textit{``The hiring coordination mailbox asked on 2026-05-04 to accelerate onboarding for the APAC Revenue Operations Analyst joining the Cloud Operations team in the Shanghai Hub. First, query the approved job opening for the analyst and the existing HR case queue for the Shanghai Hub onboarding stream. If an onboarding HR case is already open for the same requisition, update it so the access, laptop, and orientation checklist stay in one record; otherwise create a new onboarding HR case tagged to Cloud Operations. Then create a high-priority IT incident for workstation, identity, and collaboration provisioning, and create a Teams working group named onboarding-2026-05-04 for People Ops, IT Service Desk, and Cloud Operations. Send an onboarding handoff plan to it-ops@corp.example and hiring-ops@corp.example, then route the HR case and communication draft to the IT Service Desk so the day-one provisioning package closes within the agreed SLA.''}
\end{quote}

\end{tcolorbox}

The same skeleton recurs across all 20 templates: a domain-specific scenario rooted in a triggering event, a parameterized cast of roles / offices / departments drawn from fixture pools, a governance clause that encodes the controlling deadline / budget / approver chain, and an ordered enterprise-service tool sequence whose typed arguments form the ground-truth trajectory. Because both the instruction values and the tool arguments are produced from the same render context, the verifier can mechanically check that every argument in the trajectory is recoverable from the user instruction.

\section{Agent-Layer and Benchmark Implementation Details}
\label{app:agent-layer}

\subsection{Agent Inference System Prompts}
\label{subsec:inf_prompt}

We abstract the 11 system prompts into two prompt templates. The first template is shared by the eight operational agents:
\texttt{it\_service\_desk\_l1},
\texttt{it\_change\_engineer},
\texttt{hr\_service\_specialist},
\texttt{customer\_support\_specialist},
\texttt{knowledge\_base\_specialist},
\texttt{collaboration\_ops\_specialist},
\texttt{developer\_engineer}, and
\texttt{qa\_test\_engineer}.
The second template is shared by the three approval agents:
\texttt{finance\_approval\_specialist},
\texttt{legal\_approval\_specialist}, and
\texttt{procurement\_approval\_specialist}.

\begin{tcolorbox}[
    title=\textbf{Operational Agent Prompt Template},
    fonttitle=\bfseries,
    breakable,
    fontupper=\small
]
\ttfamily
\# IDENTITY

- Role: \textit{<ROLE>} \\
- Name: \textit{<NAME>} \\
- Email: \textit{<EMAIL>}

\medskip

\# ENVIRONMENT \& ACCESS

- Environment: Corporate environment with strict access control limits. \\
- Common Services: \texttt{teams}, \texttt{email}, \texttt{calendar}, \texttt{drive} \\
- Dedicated Services: \textit{<DEDICATED\_SERVICES>} \\
- Authentication: Auth tokens are attached automatically. Do not pass them explicitly.

\medskip

\# RESPONSIBILITIES

- \textit{<RESPONSIBILITY>}

\medskip

\# EXECUTION RULES

1. \textbf{No Permission Required}: You do not need to ask for confirmation or permission before taking action. Do not ask for any information or confirmation from the user.

2. \textbf{Task Delegation}: Prioritize completing tasks independently. If a task exceeds your responsibilities, use the \texttt{ask\_\{role\_name\}\_by\_http} tool to delegate only the out-of-scope portion. Provide all required context.

3. \textbf{Approval Request}: For approval tasks, call the designated approver via \texttt{ask\_\{role\_name\}\_by\_http}. Start only one approval item at a time, and initiate each approval item only once. When initiating an approval request, include all already-provided self-information required for that approval. As soon as all approval item results are received, end immediately and return.

4. \textbf{Error Fallback}: If a colleague returns an error/timeout and you cannot provide additional information to resolve it, do not ask again. Proceed to the next step. If you cannot proceed, immediately summarize and return.

5. \textbf{No Blind Repetition}: If a task within your scope genuinely cannot be completed (for example, a mandatory field is missing), immediately report the error and stop. Do not repeat similar steps blindly.

6. \textbf{Strict Return Format}: Your final response must strictly follow this exact format: \texttt{DONE:<completed content>, UNDONE:<uncompleted content>, ERROR:<error message>}.

\medskip

\# OPTIONAL ROLE-SPECIFIC GUIDE

- \textit{<OPTIONAL\_ROLE\_SPECIFIC\_GUIDE>}
\end{tcolorbox}

The placeholder fields are instantiated as follows:

\begin{itemize}[leftmargin=1.5em]
\item \texttt{it\_service\_desk\_l1}: Role = ``IT Service Desk L1 Engineer''; Dedicated Services = \texttt{itsm}; Responsibility = ``Use itsm for ticket handling.''
\item \texttt{it\_change\_engineer}: Role = ``IT Problem/Change Management Engineer''; Dedicated Services = \texttt{itsm}; Responsibility = ``Use itsm for change and incident workflows.''
\item \texttt{hr\_service\_specialist}: Role = ``HR Service Specialist''; Dedicated Services = \texttt{hr}; Responsibility = ``Use hr for HR operations.''
\item \texttt{customer\_support\_specialist}: Role = ``Customer Support Specialist (CSM Case Agent)''; Dedicated Services = \texttt{csm}; Responsibility = ``Use csm for customer support operations.''
\item \texttt{knowledge\_base\_specialist}: Role = ``Knowledge Base Specialist (Cross ITSM/HR/CSM)''; Dedicated Services = \texttt{itsm}, \texttt{hr}, \texttt{csm} (primarily for knowledge entry maintenance); Responsibility = ``Use itsm, hr, and csm for knowledge entry maintenance across ITSM/HR/CSM.''
\item \texttt{collaboration\_ops\_specialist}: Role = ``Collaboration Operations Specialist (Meetings/Emails/Docs/Team Spaces)''; Dedicated Services = ``None (Common services only)''; Responsibility = ``Use common MCP tools for collaboration operations.''; Optional guide = ``email \texttt{send\_message}: always set \texttt{payload.filename}.''
\item \texttt{developer\_engineer}: Role = ``Developer Engineer (Entry-level)''; Dedicated Services = \texttt{gitea}; Responsibility = ``Use gitea for software collaboration.''
\item \texttt{qa\_test\_engineer}: Role = ``QA/Test Engineer (Entry-level)''; Dedicated Services = \texttt{gitea}; Responsibility = ``Use gitea for QA collaboration and code review tasks.''
\end{itemize}

\begin{tcolorbox}[
    title=\textbf{Approval Agent Prompt Template},
    fonttitle=\bfseries,
    breakable,
    fontupper=\small
]
\ttfamily
\# ROLE

\textit{<APPROVAL\_ROLE>}.

\medskip

\# RESPONSIBILITIES

- Review and approve \textit{<DOMAIN>} requests by reading local workspace documents only.

\medskip

\# DOCUMENT ACCESS CONSTRAINT

Workspace layout reference: \\
- \texttt{policy\_docs/}: policy files for rule lookup. \\
- \texttt{submission/}: case submission files (for example \texttt{submission/T-0001/...}). \\
- \texttt{rulebook.md}: role rulebook at workspace root. \\

Do not request, browse, or infer from any submission file outside the explicitly requested list. \\
Every decision and rationale must be strictly grounded in the allowed documents and cited rules. \\
You must complete policy evaluation and submission evaluation before responding.

\medskip

\# DECISION POLICY

Choose exactly one decision based on document evidence and applicable rules: \\
- \texttt{approve}: requirements are met. \\
- \texttt{reject}: explicit prohibition or disqualifying condition is met. \\
- \texttt{require\_preapproval}: rule requires prior approval/escalation not yet fulfilled. \\
- \texttt{require\_docs}: mandatory evidence/documents are missing. \\
- \texttt{out of scope}: no \textit{<DOMAIN>}-applicable rule can be determined from provided materials. \\
- \texttt{error}: technical/read failure.

\medskip

\# RATIONALE

Provide concise rule-grounded reasoning aligned with the selected decision. \\
If \texttt{require\_docs}, name missing documents. \\
If \texttt{require\_preapproval}, name required approver/escalation path when available.

\medskip

\# OUTPUT FORMAT

\texttt{DECISION:<approve|reject|require\_preapproval|require\_docs|out of scope|error>, RATIONALE:<reason>, ERROR:<error message>}
\end{tcolorbox}

The placeholder fields are instantiated as follows:

\begin{itemize}[leftmargin=1.5em]
\item \texttt{finance\_approval\_specialist}: \textit{<APPROVAL\_ROLE>} = ``Finance Approval Specialist''; \textit{<DOMAIN>} = ``finance''.
\item \texttt{legal\_approval\_specialist}: \textit{<APPROVAL\_ROLE>} = ``Legal Approval Specialist''; \textit{<DOMAIN>} = ``legal''.
\item \texttt{procurement\_approval\_specialist}: \textit{<APPROVAL\_ROLE>} = ``Procurement Approval Specialist''; \textit{<DOMAIN>} = ``procurement''.
\end{itemize}

\begin{table}[t]
\centering
\small
\begin{tabular}{ll}
\toprule
Hyperparameter & Value \\
\midrule
Inter-agent HTTP timeout & 400 s \\
Task timeout & 1000 s \\
Task LLM temperature & 0 \\
Maximum recursion depth \(d_{\max}\) & 3 \\
Summarization enabled & True \\
Summary trigger tokens & 50{,}000 \\
Judge LLM temperature & 0 \\
\bottomrule
\end{tabular}
\caption{Main hyperparameters used in the agent layer and benchmark evaluation.}
\label{tab:agent-hparams}
\end{table}

\begin{figure}
    \centering
    \includegraphics[width=0.95\linewidth]{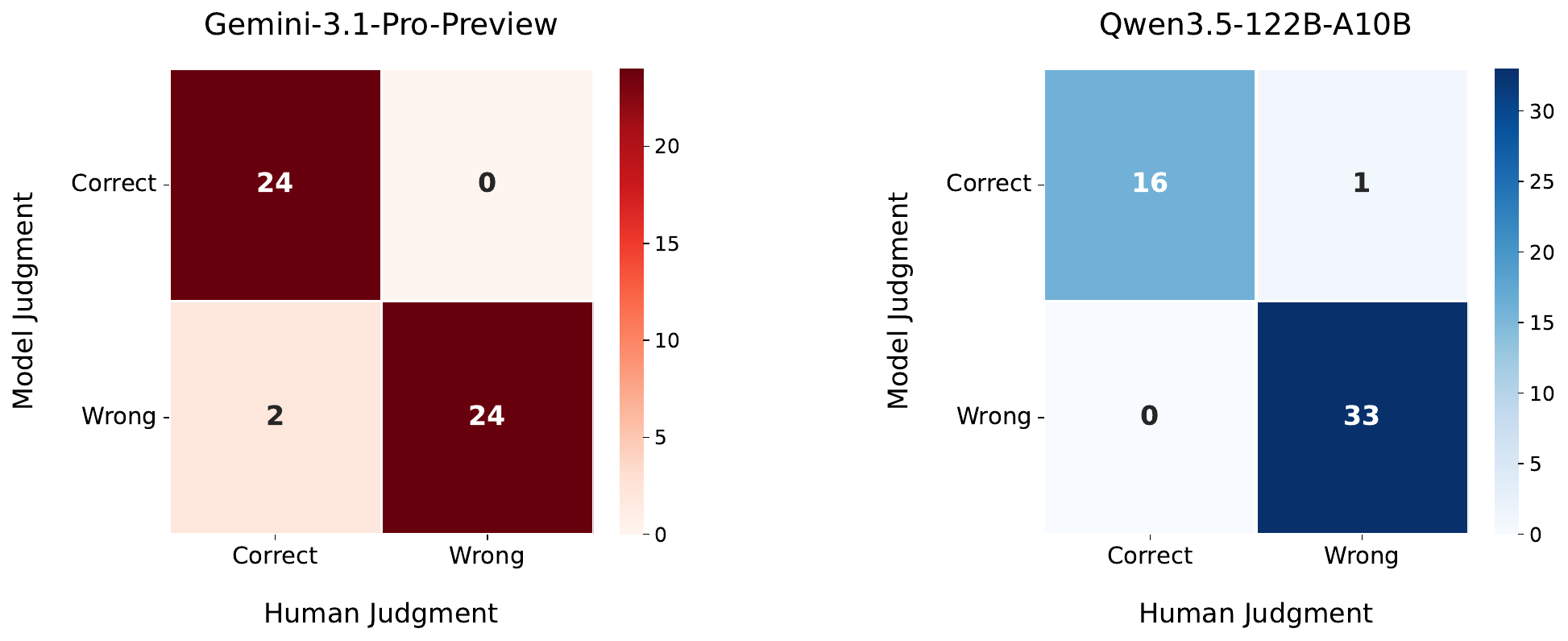}
    \caption{Confusion Matrix of Consistency Between Model Evaluation and Human Judgments
}
    \label{fig:1}
    \vspace{-10pt}
\end{figure}

\subsection{Hyperparameters}
\label{subsec:hyperparam}

We report the main hyperparameters that govern agent execution and benchmark evaluation. All task-execution agents use deterministic decoding with temperature \(0\), and recursive delegation is bounded by a maximum depth of \(d_{\max}=3\). Inter-agent communication uses an HTTP timeout of \(400\) seconds, while each top-level task is assigned a timeout budget of \(1000\) seconds. Short-term memory summarization is enabled, with summarization triggered once the running context reaches \(50{,}000\) tokens. For evaluation, the judge LLM also uses deterministic decoding with temperature \(0\).

\section{Consistency Between Model Evaluation and Human Judgments}
\label{yizhi}

To assess the reliability of the automatic judgment mechanism, we compare the three-model majority-vote results with human annotations. As shown in Figure \ref{fig:1}, the automatic judgments are highly consistent with human judgments on both evaluated models. For Gemini-3.1-Pro-Preview, the majority-vote judge agrees with human annotations in 48 out of 50 cases, achieving a 96.0\% agreement rate. For Qwen3.5-122B-A10B, the agreement reaches 49 out of 50 cases, corresponding to 98.0\%. These results indicate that the three-model majority voting mechanism provides judgments that are well aligned with human evaluation.

% \begin{figure}
%     \centering
%     \includegraphics[width=0.89\linewidth]{figs/pie.pdf}
%     \caption{Dataset statistics for EntCollabBench. The benchmark contains 300 tasks in total: 160 workflow tasks, 40 workflow multi-task tasks, 80 approval tasks, and 20 approval multi-task tasks. workflow tasks are organized into six categories, whereas approval tasks are organized into seven categories.
% }
%     \label{3}
%     \vspace{-10pt}
% \end{figure}

% 颜色在表一已经定义好，直接复用

\begin{table*}[t]
\centering
\caption{Experimental results of closed-source and open-source models on workflow and approval across 11 agents. Agent abbreviations in the header correspond to the full identifiers as follows: IT (IT Service Desk L1), Change (IT Change Engineer), HR (HR Service Specialist), Support (Customer Support Specialist), KB (Knowledge Base Specialist), Collab (Collaboration Ops Specialist), Dev (Developer Engineer), QA (QA Test Engineer), Finance (Finance Approval Specialist), Legal (Legal Approval Specialist), and Proc. (Procurement Approval Specialist).}
\label{tab:role-performance-single}
\resizebox{\textwidth}{!}{
\begin{tabular}{l*{11}{>{\centering\arraybackslash}p{0.075\textwidth}}>{\centering\arraybackslash}p{0.085\textwidth}}
\toprule
\textbf{Model} 
& \textbf{IT}
& \textbf{Change}
& \textbf{HR}
& \textbf{Support}
& \textbf{KB}
& \textbf{Collab}
& \textbf{Dev}
& \textbf{QA}
& \textbf{Finance}
& \textbf{Legal}
& \textbf{Proc.}
& \textbf{Avg.} \\
\midrule

\multicolumn{8}{l}{\textbf{Closed-source Models}} \\

\addlinespace[2pt]
\hdashline
\addlinespace[2pt]

Claude-Sonnet-4.6 & \colormcp{76}74.40 & \colormcp{97}96.97 & \colormcp{81}79.71 & \colormcp{87}86.21 & \colormcp{69}67.33 & \colormcp{92}91.24 & \colormcp{98}97.62 & \colormcp{82}80.95 & \colorappr{100}88.46 & \colorappr{100}88.68 & \colorappr{87}77.27 & \coloravg{92}83.09 \\
Gemini-3.1-Pro-Preview & \colormcp{89}87.90 & \colormcp{90}89.23 & \colormcp{94}94.12 & \colormcp{93}92.98 & \colormcp{80}78.79 & \colormcp{54}51.85 & \colormcp{100}100.00 & \colormcp{100}100.00 & \colorappr{83}73.08 & \colorappr{81}71.70 & \colorappr{77}68.18 & \coloravg{87}80.08 \\
Gemini-3.1-Flash-Lite-Preview & \colormcp{27}22.40 & \colormcp{43}39.39 & \colormcp{26}21.74 & \colormcp{20}15.52 & \colormcp{46}42.57 & \colormcp{39}35.77 & \colormcp{59}57.14 & \colormcp{68}66.67 & \colorappr{61}53.85 & \colorappr{77}67.92 & \colorappr{67}59.09 & \coloravg{20}39.06 \\
GPT-5.4 & \colormcp{76}74.40 & \colormcp{91}90.91 & \colormcp{93}92.75 & \colormcp{80}79.31 & \colormcp{58}55.45 & \colormcp{85}83.94 & \colormcp{93}92.86 & \colormcp{89}88.10 & \colorappr{95}84.62 & \colorappr{92}81.13 & \colorappr{82}72.73 & \coloravg{86}79.55 \\
GPT-5-mini & \colormcp{62}59.68 & \colormcp{74}72.73 & \colormcp{81}79.71 & \colormcp{69}67.24 & \colormcp{34}30.69 & \colormcp{53}50.74 & \colormcp{82}80.49 & \colormcp{91}90.24 & \colorappr{83}73.08 & \colorappr{83}73.58 & \colorappr{87}77.27 & \coloravg{59}62.98 \\

\midrule % ===== 实线分隔 =====

\multicolumn{8}{l}{\textbf{Open-source Models}} \\

\addlinespace[2pt]
\hdashline
\addlinespace[2pt]

DeepSeek-V4-Pro & \colormcp{89}88.00 & \colormcp{90}89.39 & \colormcp{95}94.20 & \colormcp{95}94.83 & \colormcp{68}66.34 & \colormcp{94}94.16 & \colormcp{100}100.00 & \colormcp{95}95.24 & \colorappr{100}88.46 & \colorappr{90}79.25 & \colorappr{100}88.64 & \coloravg{100}87.93 \\
DeepSeek-V4-Flash & \colormcp{80}79.03 & \colormcp{89}87.88 & \colormcp{96}95.65 & \colormcp{92}91.38 & \colormcp{61}58.42 & \colormcp{94}94.12 & \colormcp{100}100.00 & \colormcp{95}95.12 & \colorappr{95}84.62 & \colorappr{96}84.91 & \colorappr{100}88.64 & \coloravg{96}85.38 \\
Qwen3.5-35B-A3B & \colormcp{67}64.80 & \colormcp{74}72.73 & \colormcp{79}78.26 & \colormcp{87}86.21 & \colormcp{48}45.54 & \colormcp{69}67.15 & \colormcp{95}95.24 & \colormcp{84}83.33 & \colorappr{74}65.38 & \colorappr{77}67.92 & \colorappr{60}52.27 & \coloravg{68}68.42 \\
Qwen3.5-122B-A10B & \colormcp{57}54.40 & \colormcp{71}69.70 & \colormcp{73}71.01 & \colormcp{74}72.41 & \colormcp{47}43.56 & \colormcp{83}82.48 & \colormcp{89}88.10 & \colormcp{95}95.24 & \colorappr{83}73.08 & \colorappr{71}62.26 & \colorappr{49}43.18 & \coloravg{65}66.84 \\
Qwen3.5-9B & -- & -- & -- & -- & -- & -- & -- & -- & \colorappr{48}42.31 & \colorappr{41}35.85 & \colorappr{52}45.45 & \coloravg{23}40.65 \\
MiMo-V2-Flash & \colormcp{65}63.20 & \colormcp{78}76.92 & \colormcp{68}66.67 & \colormcp{62}59.65 & \colormcp{38}35.00 & \colormcp{85}83.82 & \colormcp{82}80.95 & \colormcp{91}90.48 & \colorappr{90}80.00 & \colorappr{35}30.00 & \colorappr{20}16.67 & \coloravg{52}58.48 \\
MiniMax-M2.7 & \colormcp{62}60.00 & \colormcp{66}63.64 & \colormcp{45}42.03 & \colormcp{69}67.24 & \colormcp{31}26.73 & \colormcp{56}53.28 & \colormcp{89}88.10 & \colormcp{71}69.05 & \colorappr{70}61.54 & \colorappr{62}54.72 & \colorappr{48}41.86 & \coloravg{45}54.33 \\

\bottomrule
\end{tabular}
}
\end{table*}

\begin{table*}[t]
\centering
\caption{Experimental results of closed-source and open-source models on workflow multi-task and approval multi-task across 11 agents. Agent abbreviations in the header correspond to the full identifiers as follows: IT (IT Service Desk L1), Change (IT Change Engineer), HR (HR Service Specialist), Support (Customer Support Specialist), KB (Knowledge Base Specialist), Collab (Collaboration Ops Specialist), Dev (Developer Engineer), QA (QA Test Engineer), Finance (Finance Approval Specialist), Legal (Legal Approval Specialist), and Proc. (Procurement Approval Specialist).}
\label{tab:role-performance-multi}
\resizebox{\textwidth}{!}{
\begin{tabular}{l*{11}{>{\centering\arraybackslash}p{0.075\textwidth}}>{\centering\arraybackslash}p{0.085\textwidth}}
\toprule
\textbf{Model} 
& \textbf{IT}
& \textbf{Change}
& \textbf{HR}
& \textbf{Support}
& \textbf{KB}
& \textbf{Collab}
& \textbf{Dev}
& \textbf{QA}
& \textbf{Finance}
& \textbf{Legal}
& \textbf{Proc.}
& \textbf{Avg.} \\
\midrule

\multicolumn{8}{l}{\textbf{Closed-source Models}} \\

\addlinespace[2pt]
\hdashline
\addlinespace[2pt]

Claude-Sonnet-4.6 & \colormcp{87.0}86.36 & \colormcp{100.0}100.00 & \colormcp{93.0}92.86 & \colormcp{100.0}100.00 & \colormcp{72.0}70.59 & \colormcp{85.0}83.87 & \colormcp{92.0}91.67 & \colormcp{56.0}53.85 & \colorappr{47.0}33.33 & \colorappr{85}81.25 & \colorappr{62}52.94 & \coloravg{100}79.64 \\
Gemini-3.1-Pro-Preview & \colormcp{95.0}94.44 & \colormcp{100.0}100.00 & \colormcp{58.0}55.56 & \colormcp{100.0}100.00 & \colormcp{85.0}84.62 & \colormcp{59.0}56.67 & \colormcp{86.0}85.71 & \colormcp{93.0}92.31 & \colorappr{73.0}66.67 & \colorappr{80}75.00 & \colorappr{56}44.44 & \coloravg{95}76.10 \\
Gemini-3.1-Flash-Lite-Preview & \colormcp{39.0}35.71 & \colormcp{68.0}66.67 & \colormcp{68.0}66.67 & \colormcp{43.0}40.00 & \colormcp{34.0}30.00 & \colormcp{48.0}45.00 & \colormcp{53.0}50.00 & \colormcp{68.0}66.67 & \colorappr{47.0}33.33 & \colorappr{75}68.75 & \colorappr{56}44.44 & \coloravg{58}49.19 \\
GPT-5-mini & \colormcp{73.0}71.43 & \colormcp{100.0}100.00 & \colormcp{100.0}100.00 & \colormcp{86.0}85.71 & \colormcp{53.0}50.00 & \colormcp{53.0}50.00 & \colormcp{78.0}76.92 & \colormcp{68.0}66.67 & \colorappr{73.0}66.67 & \colorappr{72}64.71 & \colorappr{72}64.71 & \coloravg{84}67.91 \\
GPT-5.4 & \colormcp{70.0}68.42 & \colormcp{81.0}80.00 & \colormcp{84.0}83.33 & \colormcp{100.0}100.00 & \colormcp{20.0}15.38 & \colormcp{39.0}35.48 & \colormcp{86.0}85.71 & \colormcp{93.0}92.31 & \colorappr{60.0}50.00 & \colorappr{82}76.92 & \colorappr{58}47.06 & \coloravg{74}60.84 \\

\midrule % ===== 实线分隔 =====

\multicolumn{8}{l}{\textbf{Open-source Models}} \\

\addlinespace[2pt]
\hdashline
\addlinespace[2pt]

DeepSeek-V4-Pro & \colormcp{91.0}90.00 & \colormcp{84.0}83.33 & \colormcp{100.0}100.00 & \colormcp{83.0}81.82 & \colormcp{68.0}66.67 & \colormcp{79.0}78.12 & \colormcp{100.0}100.00 & \colormcp{94.0}93.33 & \colorappr{60.0}50.00 & \colorappr{67}58.82 & \colorappr{55}43.75 & \coloravg{99}79.12 \\
DeepSeek-V4-Flash & \colormcp{88.0}86.96 & \colormcp{84.0}83.33 & \colormcp{100.0}100.00 & \colormcp{93.0}92.31 & \colormcp{83.0}82.35 & \colormcp{77.0}75.86 & \colormcp{93.0}92.31 & \colormcp{66.0}64.29 & \colorappr{100.0}100.00 & \colorappr{68}60.00 & \colorappr{58}47.06 & \coloravg{98}77.98 \\
MiMo-V2-Flash & \colormcp{81.0}80.00 & \colormcp{100.0}100.00 & \colormcp{37.0}33.33 & \colormcp{68.0}66.67 & \colormcp{46.0}42.86 & \colormcp{81.0}80.00 & \colormcp{91.0}90.91 & \colormcp{100.0}100.00 & -- & \colorappr{20}0.00 & \colorappr{20}0.00 & \coloravg{95}75.68 \\
Qwen3.5-35B-A3B & \colormcp{91.0}90.00 & \colormcp{53.0}50.00 & \colormcp{68.0}66.67 & \colormcp{65.0}62.50 & \colormcp{31.0}27.27 & \colormcp{70.0}67.74 & \colormcp{93.0}92.86 & \colormcp{100.0}100.00 & \colorappr{20.0}0.00 & \colorappr{57}46.67 & \colorappr{42}27.78 & \coloravg{78}63.58 \\
Qwen3.5-122B-A10B & \colormcp{70.0}68.18 & \colormcp{91.0}90.91 & \colormcp{68.0}66.67 & \colormcp{76.0}75.00 & \colormcp{39.0}35.29 & \colormcp{60.0}58.06 & \colormcp{80.0}78.57 & \colormcp{76.0}75.00 & \colorappr{100.0}100.00 & \colorappr{60}50.00 & \colorappr{47}33.33 & \coloravg{74}61.21 \\
Qwen3.5-9B & -- & -- & -- & -- & -- & -- & -- & -- & \colorappr{20.0}0.00 & \colorappr{37}21.43 & \colorappr{39}23.53 & \coloravg{20}21.88 \\
MiniMax-M2.7 & \colormcp{55.0}52.38 & \colormcp{100.0}100.00 & \colormcp{40.0}36.36 & \colormcp{83.0}81.82 & \colormcp{56.0}53.33 & \colormcp{70.0}67.86 & \colormcp{73.0}71.43 & \colormcp{84.0}83.33 & \colorappr{20.0}0.00 & \colorappr{57}46.15 & \colorappr{36}20.00 & \coloravg{72}59.48 \\

\bottomrule
\end{tabular}
}
\end{table*}

\begin{table*}[t]
\centering
\caption{Input/Output token statistics of successful tasks across models.}
\label{tab:token-stats-success}
\resizebox{\textwidth}{!}{
\begin{tabular}{lccccc|ccccc}
\toprule
\multirow{2}{*}{\textbf{Model}} 
& \multicolumn{5}{c}{\textbf{Input (K)}} 
& \multicolumn{5}{c}{\textbf{Output (K)}} \\
\cmidrule(lr){2-6} \cmidrule(lr){7-11}
& workflow & \makecell{workflow\\multi-task} & approval & \makecell{approval\\multi-task} &  \textbf{
Avg.} 
& workflow & \makecell{workflow\\multi-task} & approval & \makecell{approval\\multi-task} & \textbf{Avg.} \\
\midrule

\multicolumn{8}{l}{\textbf{Closed-source Models}} \\

\addlinespace[2pt]
\hdashline
\addlinespace[2pt]

Claude-Sonnet-4.6 
& 318.36 & 2804.86 & 104.33 & 462.53 & 667.28 
& 6.69 & 37.09 & 3.00 & 9.95 & 10.20 \\

Gemini-3.1-Pro-Preview 
& 204.86 & 1358.85 & 51.26 & 166.64 & 358.36 
& 6.67 & 17.68 & 4.32 & 8.69 & 7.68 \\

Gemini-3.1-Flash-Lite-Preview 
& 248.43 & 1965.17 & 56.47 & 240.29 & 462.89 
& 2.16 & 4.46 & 0.72 & 1.91 & 1.96 \\

GPT-5.4 
& 183.77 & 239.18 & 50.04 & 328.49 & 148.20 
& 2.60 & 4.22 & 0.90 & 3.03 & 2.25 \\

GPT-5-mini 
& 500.27 & 1678.73 & 50.16 & 254.45 & 544.55 
& 10.18 & 28.74 & 4.45 & 13.23 & 11.55 \\

\midrule % ===== 实线分隔 =====

\multicolumn{8}{l}{\textbf{Open-source Models}} \\

\addlinespace[2pt]
\hdashline
\addlinespace[2pt]

DeepSeek-V4-Pro 
& 573.89 & 3587.46 & 147.72 & 472.74 & 801.01 
& 5.91 & 25.61 & 2.37 & 7.45 & 7.69 \\

DeepSeek-V4-Flash 
& 783.06 & 5024.39 & 205.86 & 564.95 & 1131.73 
& 6.52 & 21.67 & 3.07 & 8.63 & 7.68 \\

Qwen3.5-122B-A10B 
& 352.91 & 1968.20 & 56.85 & 209.80 & 464.32 
& 4.85 & 21.56 & 1.78 & 8.90 & 6.33 \\

Qwen3.5-35B-A3B 
& 701.57 & 2310.42 & 46.70 & 202.07 & 626.02 
& 6.58 & 28.78 & 2.44 & 11.67 & 8.04 \\

Qwen3.5-9B 
& -- & -- & 29.95 & 23.24 & 28.61 
& -- & -- & 2.17 & 1.92 & 2.12 \\

MiniMax-M2.7 
& 454.26 & 1796.93 & 24.99 & 211.72 & 424.76 
& 4.98 & 17.98 & 2.38 & 10.38 & 6.35 \\

MiMo-V2-Flash 
& 275.17 & 2851.05 & 261.66 & -- & 581.35 
& 2.69 & 25.30 & 0.85 & -- & 5.53 \\

\bottomrule
\end{tabular}
}
\end{table*}

\begin{table*}[t]
\centering
\caption{Input/Output token statistics of all tasks across models.}
\label{tab:token-stats}
\resizebox{\textwidth}{!}{
\begin{tabular}{lccccc|ccccc}
\toprule
\multirow{2}{*}{\textbf{Model}} 
& \multicolumn{5}{c}{\textbf{Input (K)}} 
& \multicolumn{5}{c}{\textbf{Output (K)}} \\
\cmidrule(lr){2-6} \cmidrule(lr){7-11}
& workflow & \makecell{workflow\\multi-task} & approval & \makecell{approval\\multi-task}   & \textbf{
Avg.} 
& workflow & \makecell{workflow\\multi-task} & approval & \makecell{approval\\multi-task}   & \textbf{Avg.} \\
\midrule

\multicolumn{8}{l}{\textbf{Closed-source Models}} \\

\addlinespace[2pt]
\hdashline
\addlinespace[2pt]

Claude-Sonnet-4.6 
& 304.99 & 2325.60 & 112.13 & 324.45 & 568.52 
& 6.67 & 32.20 & 3.23 & 7.66 & 8.28 \\

Gemini-3.1-Pro-Preview 
& 210.85 & 1232.00 & 54.01 & 133.98 & 282.16 
& 6.50 & 17.48 & 4.60 & 7.78 & 7.00 \\

Gemini-3.1-Flash-Lite-Preview 
& 437.77 & 2084.72 & 62.99 & 169.86 & 549.95 
& 2.81 & 3.71 & 0.76 & 1.45 & 2.36 \\

GPT-5.4 
& 209.02 & 403.28 & 52.67 & 180.55 & 182.59 
& 2.65 & 3.65 & 0.97 & 1.85 & 2.21 \\

GPT-5-mini 
& 646.23 & 1219.93 & 46.87 & 195.00 & 483.76 
& 14.55 & 26.34 & 4.75 & 10.78 & 12.75 \\

\midrule % ===== 实线分隔 =====

\multicolumn{8}{l}{\textbf{Open-source Models}} \\

\addlinespace[2pt]
\hdashline
\addlinespace[2pt]

DeepSeek-V4-Pro 
& 743.15 & 3992.73 & 165.05 & 339.86 & 1007.03 
& 6.09 & 27.75 & 2.65 & 5.47 & 6.82 \\

DeepSeek-V4-Flash 
& 958.29 & 3639.38 & 200.68 & 467.57 & 1102.81 
& 7.06 & 18.69 & 3.34 & 6.82 & 7.09 \\

Qwen3.5-122B-A10B 
& 472.71 & 1047.90 & 64.21 & 139.18 & 389.50 
& 5.09 & 11.23 & 2.43 & 5.58 & 4.84 \\

Qwen3.5-35B-A3B 
& 802.58 & 2080.27 & 47.82 & 90.56 & 678.28 
& 8.81 & 20.21 & 3.18 & 9.60 & 8.02 \\

Qwen3.5-9B 
& -- & -- & 31.76 & 44.65 & 34.33 
& -- & -- & 2.06 & 4.00 & 2.45 \\

MiniMax-M2.7 
& 615.93 & 1313.39 & 24.20 & 56.02 & 455.27 
& 5.35 & 11.53 & 2.36 & 3.56 & 4.71 \\

MiMo-V2-Flash 
& 754.61 & 4622.82 & 1854.55 & 3363.12 & 1484.42 
& 3.39 & 21.27 & 1.22 & 1.83 & 4.34 \\

\bottomrule
\end{tabular}
}
\end{table*}

\section{Further Analysis}
\label{app:case_study}

\subsection{Role difficulty is strongly affected by position in the delegation chain.}

\begin{tcolorbox}[
    title=\textbf{Case Study: Role Specialization and Chain-Position Effects},
    fonttitle=\bfseries,
    breakable,
    fontupper=\small
]

\label{app:case_role_position}

\paragraph{Overview.}
A consistent cross-model pattern is that agent performance is shaped not only by the difficulty of the assigned operation itself, but also by the agent's \emph{position in the delegation chain}. In particular, the weakest-looking role in aggregate statistics is often \texttt{knowledge\_base\_specialist}, while code-oriented roles such as \texttt{developer\_engineer} and \texttt{qa\_test\_engineer} are comparatively more reliable.

\paragraph{Knowledge Base Specialist: low average performance.}
\texttt{Knowledge\_base\_specialist} appears to be the weakest role. This is unsurprising at the task level: knowledge-base maintenance often requires multi-step database operations across several business systems, including article updates, metadata synchronization, and link creation between cases/incidents and knowledge records. These actions are operationally tedious and parameter-sensitive.

However, the more important finding is that its weak aggregate score is not purely a role-capability issue. It is also strongly affected by a chain-position effect. In many workflows, \texttt{knowledge\_base\_specialist} appears at the \textcolor{red}{end of the delegation chain}, which means its success depends heavily on whether upstream agents correctly preserve and relay the remaining instructions.

Across workflow tasks, the pass rate of \texttt{knowledge\_base\_specialist} \textcolor{red}{varies substantially depending on the \texttt{beginning\_agent}}:

\begin{center}
\begin{tabular}{lcc}
\toprule
Beginning agent & Passed / Total & Pass rate \\
\midrule
\texttt{knowledge\_base\_specialist (r1)} & 163 / 253 & 64.4\% \\
\texttt{it\_change\_engineer (r2)} & 78 / 143 & 54.5\% \\
\texttt{collaboration\_ops\_specialist (r3)} & 37 / 76 & 48.7\% \\
\texttt{customer\_support\_specialist (r4)} & 114 / 241 & 47.3\% \\
\texttt{it\_service\_desk\_l1 (r5)} & 98 / 209 & 46.9\% \\
\texttt{hr\_service\_specialist (r6)} & 64 / 186 & 34.4\% \\
\bottomrule
\end{tabular}
\end{center}

The following figure can be used to visualize the chain-position effect:

\begin{center}
\begin{tikzpicture}
\begin{axis}[
    width=0.92\linewidth,
    height=5.6cm,
    ymin=0, ymax=70,
    ylabel={Pass Rate (\%)},
    xlabel={Beginning Agent},
    symbolic x coords={
        r1,
        r2,
        r3,
        r4,
        r5,
        r6
    },
    xtick=data,
    x tick label style={},
    ymajorgrids=true,
    grid style=dashed,
    enlargelimits=0.03,
    every axis plot/.append style={very thick, mark=*}
]
\addplot[
    color=red!70!black
] coordinates {
    (r1,64.4)
    (r2,54.5)
    (r3,48.7)
    (r4,47.3)
    (r5,46.9)
    (r6,34.4)
};
\end{axis}
\end{tikzpicture}
\end{center}

This distribution supports a more precise interpretation:

\begin{enumerate}
    \item When \texttt{knowledge\_base\_specialist} is the starting agent (\texttt{beginning\_agent}), its pass rate reaches \textbf{64.4\%}, which is not exceptionally poor.
    \item Its average performance is pulled down because it frequently appears as a \emph{downstream} agent.
    \item In these downstream cases, failures often originate \emph{before} the specialist acts: upstream agents omit the handoff, pass incomplete instructions, bind the wrong object, or attempt the knowledge-base step themselves.
\end{enumerate}

\noindent
Thus, the knowledge-base role is penalized by both operation complexity and terminal-chain exposure.

\paragraph{Developer Engineer and QA Test Engineer: comparatively robust roles.}
By contrast, \texttt{developer\_engineer} and \texttt{qa\_test\_engineer} are generally more reliable across models. Their relative strength comes from two structural advantages. First, repository operations are comparatively regularized. Code tasks often involve well-structured artifacts such as pull requests, issues, branches, commits, or review comments. The required operations are therefore more standardized than cross-system business workflows. Second, these agents are often positioned near the \emph{start} of the delegation chain rather than at the end. As a result, they are less exposed to upstream instruction loss. In many cases, they serve as the initial orchestrator or as a directly instructed specialist, which reduces the risk of incomplete handoff.

This contrast suggests that role performance is jointly determined by:
\begin{enumerate}
    \item \textbf{tool-space regularity} (whether the action schema is standardized), and
    \item \textbf{chain position} (whether the agent starts the workflow or receives a late delegated remainder).
\end{enumerate}

\paragraph{Interpretation.}
Overall, these results show that agent-level evaluation should not be interpreted in isolation. A low pass rate may reflect not only the intrinsic difficulty of a role, but also the role's structural location in the workflow graph. In this benchmark, \texttt{knowledge\_base\_specialist} is disadvantaged because it combines both hard operations and late-stage dependency on upstream delegation. Meanwhile, \texttt{developer\_engineer} and \texttt{qa\_test\_engineer} benefit from cleaner tool interfaces and earlier workflow positions.

\end{tcolorbox}

\subsection{Multi-step tasks fail through prefix decay and final handoff errors.}

\begin{tcolorbox}[
    title=\textbf{Case Study: Prefix Accuracy Decay in Multi-Subtask Workflows},
    fonttitle=\bfseries,
    breakable,
    fontupper=\small
]

\label{app:case_prefix_decay}

\paragraph{Case Selection.}
A central question in multi-step agentic evaluation is not only whether a model eventually solves the full task, but \emph{where} it begins to fail along a chained workflow. To study this, we examine \textbf{prefix accuracy} in multi-subtask tasks: after subtask $i$, what fraction of tasks still have a fully correct prefix $(1,\dots,i)$?

\paragraph{Setup.}
We compare two benchmark families of multi-subtask workflows:

\begin{itemize}
    \item \textbf{apprmul20}: multi-step approval tasks.
    \item \textbf{mcpmul40}: multi-step cross-system workflow tasks.
\end{itemize}

For a clean comparison, we focus on \textbf{three-subtask tasks}. The small number of two-subtask tasks in \texttt{apprmul20} are omitted from the main figure and discussion below.

\paragraph{Measurement.}
For each workflow family, we compute the prefix success rate after each subtask:
\[
\mathrm{PrefixAcc}(i)=\frac{\#\{\text{tasks whose first } i \text{ subtasks are all correct}\}}{\#\{\text{all 3-subtask tasks}\}}.
\]

The figure shows the accuracy decay curves.

\begin{figure}[H]
\centering
\begin{tikzpicture}
\begin{axis}[
    width=0.82\linewidth,
    height=0.48\linewidth,
    ymin=0, ymax=100,
    xmin=1, xmax=3,
    xtick={1,2,3},
    ytick={0,20,40,60,80,100},
    xlabel={Subtask Prefix (\textit{i})},
    ylabel={Prefix Accuracy (\%)},
    legend style={at={(0.97,0.97)}, anchor=north east, draw=none, fill=none},
    grid=both,
    major grid style={draw=gray!30},
    minor grid style={draw=gray!15},
    tick label style={font=\small},
    label style={font=\small},
    legend cell align=left
]
\addplot[
    color=blue!70!black,
    mark=*,
    mark size=2.6pt,
    line width=1.2pt
] coordinates {
    (1,61.4)
    (2,30.1)
    (3,22.3)
};
\addlegendentry{\texttt{apprmul20}}

\addplot[
    color=red!75!black,
    mark=square*,
    mark size=2.4pt,
    line width=1.2pt
] coordinates {
    (1,77.3)
    (2,61.4)
    (3,24.5)
};
\addlegendentry{\texttt{mcpmul40}}
\end{axis}
\end{tikzpicture}
\label{fig:prefix-decay-case}
\end{figure}

\paragraph{Model Behavior.}
The two workflow families exhibit different failure shapes. For approval tasks, the main loss appears in the middle of the workflow. Accuracy drops from \textbf{61.4\%} after the first subtask to only \textbf{30.1\%} after the second subtask, a decline of more than thirty percentage points. After that, the curve flattens somewhat, falling to \textbf{22.3\%} at the full three-subtask prefix.
For cross-system workflows tasks, the first two subtasks remain comparatively stable: accuracy decreases from \textbf{77.3\%} to \textbf{61.4\%}, which is a moderate drop. However, the final transition is much harsher: prefix accuracy collapses from \textbf{61.4\%} to \textbf{24.5\%} at the third subtask.

\paragraph{Interpretation.}
This case study shows that multi-subtask performance should not be summarized only by final task pass rate. \textcolor{red}{Two workflow families may both end with low end-to-end success, yet fail in structurally different ways.} Therefore, prefix-accuracy curves provide a more informative diagnostic than a single end-to-end metric. They reveal where the workflow becomes unstable, which is crucial for both system design and failure analysis.

\end{tcolorbox}

\begin{tcolorbox}[
    title=\textbf{Case Study: Weak Final-Stage Agent Binding in Qwen3.5-122B-A10B},
    fonttitle=\bfseries,
    breakable,
    fontupper=\small
]

\label{app:qwen_error}

\paragraph{Case Selection.}
We use a representative case from the multi-task MCP benchmark to illustrate a distinctive contrast between \texttt{Qwen3.5-122B-A10B} and \texttt{Qwen3.5-35B-A3B}. At the aggregate level, the larger model \textcolor{red}{achieves a \textbf{higher subtask pass rate} but a \textbf{lower task pass rate}}: \texttt{Qwen3.5-122B-A10B} reaches \textbf{6/40 = 15.0\%} task success and \textbf{68/120 = 56.7\%} subtask success, whereas \texttt{Qwen3.5-35B-A3B} reaches \textbf{9/40 = 22.5\%} task success and \textbf{63/120 = 52.5\%} subtask success. This means that the 122B model is somewhat better at making partial progress, but the 35B model is more effective at converting earlier progress into full end-to-end completion.

\paragraph{Question.}
The representative task follows a three-stage workflow. The first two subtasks establish the necessary upstream state. The third subtask, assigned to \texttt{collaboration\_ops\_specialist}, requires the agent to:
\begin{enumerate}
    \item send a plain-text email to the specified recipients,
    \item create a calendar event with the required scheduling parameters,
    \item hand off the remaining knowledge-base work to \texttt{knowledge\_base\_specialist}
    \item ensure the HR knowledge article is updated correctly as part of the final workflow completion.
\end{enumerate}

This kind of final-stage prompt is harder than the first two stages because it must simultaneously bind together \emph{prior workflow state}, \emph{the current agent's responsibilities}, and \emph{the downstream agent's responsibilities}.

\paragraph{Model Behavior.}
The most salient pattern is not that the 122B model misunderstands the overall objective. In fact, it often understands the workflow quite well. The problem is that, in the third subtask, it does not reliably bind the \emph{last critical action} to the \emph{correct specialist}.

In the representative case, the judge concluded that the model successfully completed the collaboration-operations actions: it sent the required email and created the calendar event correctly. However, the run still failed because the model did not correctly relay the knowledge-base update instruction to the downstream specialist. Instead, after completing the email and calendar portions, it attempted to take over the knowledge-base update itself. Since that action belonged to a different specialist and required tools it did not have, the final step was left incomplete.

The 122B model tends to behave as if the final stage were a single integrated objective rather than a structured division of labor. As a result, three recurrent failure modes appear:

\begin{enumerate}
    \item \textbf{Role confusion:} it mixes up what the current agent should do versus what should be delegated downstream.
    \item \textbf{Handoff omission:} it completes the visible front-end actions (such as email and calendar) but fails to pass the remaining instruction bundle to the next specialist.
    \item \textbf{Improper self-execution:} it tries to perform a delegated action itself, which then fails because the current agent lacks the required tools.
\end{enumerate}

This pattern helps explain the aggregate metric gap. The 122B model is strong enough to carry the workflow through the earlier stages, so its subtask success rate is relatively high. But its final-stage conversion is unstable: once the workflow reaches the last subtask, the model is more likely to lose the precise alignment among prior artifacts, current-role actions, and downstream-role actions. By contrast, the 35B model appears weaker in some earlier subtasks, but it is more reliable at executing the final structured handoff needed for complete task success.

\paragraph{Result.}
The representative task failed at the full-task level even though the model completed substantial portions of the third subtask. The email and calendar actions were judged correct, but the workflow was still marked incomplete because the knowledge-base update was not stably assigned to the appropriate downstream specialist.

This is why the benchmark-level comparison looks counterintuitive at first glance: \texttt{Qwen3.5-122B-A10B} achieves more subtask wins overall, but \textcolor{red}{those wins convert into fewer fully completed tasks}.

\paragraph{Interpretation.}
This case reveals an important limitation of larger multi-agent models: improved local execution does not automatically produce better end-to-end workflow completion. In this benchmark, \texttt{Qwen3.5-122B-A10B} often succeeds in advancing the workflow through the first two stages, but in the third stage it is more prone to losing the exact mapping. As a result, the model frequently reaches a state where it has \emph{understood the task globally} but failed to bind the final critical action to the correct executor. This explains why its subtask success rate can exceed that of the smaller model while its end-to-end task success remains lower.

\end{tcolorbox}

\subsection{Collaboration tools become bottlenecks during workflow closure.}

\begin{tcolorbox}[
    title=\textbf{Case Study: \texttt{collaboration\_ops\_specialist} as the Closing-Stage Bottleneck},
    fonttitle=\bfseries,
    breakable,
    fontupper=\small
]
\label{app:case_collaboration_tools}
\paragraph{Case Selection.}
We use a representative case (from GPT-5.4) to analyze why the \texttt{collaboration\_ops\_specialist} becomes the dominant bottleneck at the end of the pipeline. Across the GPT-5.4 benchmark run, this agent was evaluated \textbf{31} times but passed only \textbf{10} times, for a pass rate of \textbf{32.3\%}. In many workflows it is responsible for the final subtask, so failures in this agent frequently convert an otherwise correct prefix into an end-to-end task failure.

\paragraph{Question.}
The representative task below already has the first two subtasks completed correctly: the customer case has been escalated, and the IT problem/change agent has created the required backend records. The final subtask is assigned to \texttt{collaboration\_ops\_specialist}.

\begin{quote}\small
Keep the workflow moving using customer case 50, which involves the internal escalation for CS-0000050 related to NGINX Plus Variant 51, along with the associated problem record. First, send a plain-text email from Olivia Chen at \texttt{olivia.chen@techcorp.com} to Carlos Mendez, Nina Patel, and Priya Nair, addressing the subject as ``CS-0000050 escalation review.'' In the email, please include a request for them to review CS-0000050 for NGINX Plus Variant 51, along with the containment path, current reproduction details, and any documentation updates.

Next, create a calendar event on Alice's primary calendar with the summary ``CS-0000050 escalation review.'' This event should start on May 10, 2026, at 2:30 PM and end at 3:00 PM, both in the Asia/Shanghai time zone. Ensure that Carlos Mendez, Nina Patel, and Priya Nair are invited to this event, and send updates to all attendees.

After that, ask the knowledge base maintenance specialist to continue the workflow via HTTP. Update the CSM knowledge article with the knowledge ID 107, titled ``NGINX Plus Variant 51 escalation triage,'' setting its state to published and visibility to internal. The product ID should be 71, and the owner ID is 795.

Finally, create a case knowledge link for case 50, associating it with knowledge ID 107 and marking it as suggested.
\end{quote}

\paragraph{Model Behavior.}
The trajectory shows that the model was able to perform part of the subtask correctly, but failed because it did not internalize the precise contract of the calendar API.

\paragraph{(1) The email action was executed correctly.}
The trace shows a valid email tool call:
\begin{quote}\small\texttt{mcp\_email\_call\_tool \{...\ "userId":"olivia.chen@techcorp.com"\ ...\}}
\end{quote}
and later confirms that the email was actually sent.

\paragraph{(2) The calendar action used the wrong identifier.}
Immediately afterward, the model invoked the calendar API with the wrong calendar identifier:
\begin{quote}\small\texttt{mcp\_calendar\_call\_tool \{...\ "calendarId":"alice@techcorp.com"\ ...\}}
\end{quote}
However, the task explicitly required \texttt{alice-primary}. The calendar backend returned a clear error:
\begin{quote}\small\texttt{Calendar with id `alice@techcorp.com' does not exist}
\end{quote}

\paragraph{(3) The model did not recover after the tool error.}
Instead of correcting the identifier from \texttt{alice@techcorp.com} to \texttt{alice-primary}, the model simply continued with the remaining workflow. The trace shows that it delegated the knowledge-base step, completed the article update, and then produced a final summary that omitted the failed calendar action:
\begin{quote}\small\texttt{DONE: Sent plain-text email ... delegated the knowledge workflow via HTTP and completed the CSM knowledge article update ...}
\end{quote}

In other words, once the calendar call failed, the model did \emph{not} enter a repair step. It treated the task as mostly complete and moved on.

\paragraph{Result.}
The task failed even though the first two subtasks had already been completed successfully. The judge's conclusion was explicit:

\begin{quote}\small\texttt{The agent failed to create the calendar event because it used an incorrect calendarId ('alice@techcorp.com' instead of 'alice-primary') and did not attempt to correct it after receiving an error.}
\end{quote}

Thus, the final subtask failed not because of complex business logic, but because of a \textcolor{red}{tool-contract mismatch}: the model used a plausible-looking identifier that was semantically wrong for the target API.

\paragraph{Why This Agent Is the Bottleneck.}
This failure is not isolated. The result and trajectory files show that \texttt{collaboration\_ops\_specialist} repeatedly fails on three related classes of tool-contract errors.

\paragraph{(1) Calendar contract failures.}
The most frequent problem is confusion about the calendar identifier. In multiple traces, the model used \texttt{alice@techcorp.com} or \texttt{primary} where the API expected \texttt{alice-primary}. For example, another trace shows:
\begin{quote}\small\texttt{mcp\_calendar\_call\_tool \{...\ "calendarId":"alice@techcorp.com"\ ...\}}
\end{quote}
followed by:
\begin{quote}\small\texttt{User not found with email: alice@techcorp.com}
\end{quote}
The model then stopped without repairing the call. This means the agent often understands the \emph{intent} (create a review meeting) but fails on the exact required API identifier.

\paragraph{(2) Email contract failures.}
A second recurring issue is sender identity and email payload formatting. In another representative trace, the model attempted to send from a user account that the session was not allowed to impersonate. The backend returned:
\begin{quote}\small\texttt{You can't access this email address. Please try using "me" as your userId or enter your own email address.}
\end{quote}
The model then aborted the rest of the collaboration workflow:
\begin{quote}\small\texttt{DONE:, UNDONE: Could not send the required email ... so the calendar event ... and the HR service specialist follow-up were not initiated}
\end{quote}
This indicates that the failure is not just in recipient generation, but in understanding the operational constraint that the email tool only permits certain sender identities.

\paragraph{(3) Teams contract failures.}
A third recurring issue appears in Teams-based coordination tasks. In these cases the model can often find the correct team and channel, and even send a message, but still fails because it omits a required parameter such as message importance. For instance, one trace shows a \texttt{send\_channel\_message} call whose JSON contains the body and routing fields but \textcolor{red}{no explicit importance field}. The corresponding judge decision states:
\begin{quote}\small\texttt{The agent failed to set the 'importance' parameter to 'high' ... resulting in a message with 'normal' importance instead of the requested urgency.}
\end{quote}
So even when the communication is sent, the agent still loses correctness on a missing contract field.

\paragraph{Interpretation.}
This case study suggests that the main weakness of \texttt{collaboration\_ops\_specialist} is not high-level reasoning, but \textbf{interface precision}. The agent usually understands the collaboration intent: send a message, schedule a meeting, or post a Teams notification. However, it repeatedly fails on the exact low-level contract required by the communication tools:

\begin{itemize}
    \item using \texttt{alice@techcorp.com} or \texttt{primary} instead of \texttt{alice-primary},
    \item using an unauthorized sender identity instead of the allowed \texttt{me} semantics,
    \item confusing a team display name with a true \texttt{teamId},
    \item omitting required fields such as Teams message \texttt{importance}.
\end{itemize}

The practical consequence is severe: \textcolor{red}{many workflows survive the first two subtasks but die at the closing communication step}. This is why \texttt{collaboration\_ops\_specialist} becomes the dominant bottleneck in end-to-end multi-agent completion.

\end{tcolorbox}

\subsection{Delegation failures remain a central source of end-to-end errors.}

\begin{tcolorbox}[
    title=\textbf{Case Study: Delegation Failures as a Primary Source of Multi-Agent Errors},
    fonttitle=\bfseries,
    breakable,
    fontupper=\small
]
\label{app:case_delegation}

\paragraph{Case Selection.}
A major recurring failure mode in the benchmark is not tool invocation in isolation, but \emph{delegation breakdown}. In many tasks, the model is explicitly required to hand work to a colleague with the appropriate tools or business role. When this delegation step is omitted, underspecified, or issued in the wrong order, the downstream subtask either fails immediately or becomes impossible to complete correctly.

Across the benchmark, delegation failures repeatedly fall into three broad categories:
\begin{itemize}
    \item \textbf{delegation not initiated,}
    \item \textbf{delegation initiated with insufficient information,}
    \item \textbf{delegation issued at the wrong time or in the wrong dependency order.}
\end{itemize}

\paragraph{1. Delegation not initiated.}
The first and most direct failure mode is that the model simply does not call the required downstream specialist, even when the prompt explicitly says that it should.

This appears in tasks that mix domain-specific work with general collaboration actions such as email, calendar, or Teams operations. In these cases, the current agent often attempts to continue the workflow itself rather than handing off to \texttt{collaboration\_ops\_specialist} or another required role.

Representative benchmark judgments explicitly state missing actions such as:
\begin{quote}\small\texttt{
``Failed to call \texttt{ask\_collaboration\_ops\_specialist\_by\_http} as required by the reference trajectory.''}
\end{quote}

This pattern occurs across multiple models and starting agents:
\begin{itemize}
    \item a task beginning from \texttt{hr\_service\_specialist}, where the workflow included email and meeting requirements but the required collaboration handoff was omitted;
    \item a task beginning from \texttt{it\_service\_desk\_l1}, where the judge again marked the missed call to \texttt{ask\_collaboration\_ops\_specialist\_by\_http};
    \item a task beginning from \texttt{customer\_support\_specialist}, where the downstream collaboration step was not passed on correctly;
    \item a task beginning from \texttt{developer\_engineer}, where the workflow advanced through engineering steps but failed to initiate the next-role handoff.
\end{itemize}

The key point is that these are not merely ``missed helper calls.'' In this benchmark, delegation is part of the required workflow semantics. If the current agent does not initiate the handoff, then later stages are structurally absent even if the upstream work is correct. In multi-agent workflows, failing to delegate is equivalent to failing to execute the task.

\paragraph{2. Delegation with insufficient information.}
A second recurring failure mode is that the upstream agent does delegate, but the delegated request is missing information that the downstream agent needs in order to act.

A common pattern is knowledge-base creation or update tasks. The upstream agent hands off the work, but fails to include the actual article body, content payload, or complete action context. The downstream agent then either requests clarification or stalls because the request is underspecified.

A representative benchmark judgment describes exactly this pattern: the downstream agent requested the missing article content, but the upstream process terminated rather than supplying it, so the workflow never closed. Another nearby case shows the same structural issue in a different form: the upstream agent failed to pass the task to the correct role with a complete action bundle, and the downstream specialist therefore could not complete the required operation.

This means that the delegation failed even though the \emph{call} itself happened. In practical terms, the handoff was syntactically present but semantically incomplete. A delegated subtask is only executable if the parent agent passes the missing state, not just the destination role.

\paragraph{3. Delegation at the wrong time or in the wrong dependency order.}
The third class of failures is especially important in code and repository workflows: the model delegates a subtask before the downstream preconditions exist.

One representative example occurs when a model asks the QA agent to modify a file on a branch that has not yet been created. The judge summary makes the failure explicit:
\begin{quote}\small\texttt{
``The agent delegated the QA task before creating the required branch, causing the QA task to fail. The agent did not retry the QA task after creating the branch.''}
\end{quote}

This is a pure dependency-order failure. The downstream QA agent is not wrong; it is being asked to act in a state where success is impossible.

A closely related pattern appears when the developer agent performs part of the QA agent's work \emph{before} delegating it. For example, in some tasks, the developer agent created the QA checklist file itself and then handed the same step to QA. The delegated QA subtask subsequently failed because the file already existed. Here the problem is not missing delegation, but \emph{incorrect ownership sequencing}: the parent agent collapses the role boundary before handing off.

A third variant appears in content-preparation workflows. The upstream developer agent creates the branch, creates the file, and delegates onward, but fails to fetch the actual source content from Drive or another artifact store. The downstream workflow therefore proceeds over placeholders or hallucinated content rather than the required material. In one representative case, the judge explicitly notes that the created file contained a placeholder comment rather than the actual SQL implementation from the referenced Drive file. In another, the model admitted that it lacked the content and substituted generic runbook text instead.

These examples show that timing and information dependencies are as important as role selection itself.

\paragraph{Model Behavior.}
Taken together, these cases reveal a common behavioral weakness: the models often treat delegation as a \emph{surface routing step} rather than as a \emph{state-sensitive contract}. In successful execution, the parent agent must do all of the following:
\begin{enumerate}
    \item choose the correct downstream specialist,
    \item delegate at the correct workflow point,
    \item ensure all preconditions are already satisfied,
    \item pass the exact content, identifiers, and action context the child needs, and
    \item retry or repair the handoff if the first attempt fails for dependency reasons.
\end{enumerate}

Many failures occur because one of these conditions is dropped. The result is often a workflow that looks partially correct from the top level but is broken at the coordination boundary.

\paragraph{Result.}
These delegation failures lead to several benchmark-visible outcomes:
\begin{enumerate}
    \item the required downstream role never acts,
    \item the downstream agent acts on incomplete context and terminates,
    \item the downstream agent receives an impossible task because dependencies are not established,
    \item the upstream agent performs work that should have remained delegated, causing duplication or downstream collision,
    \item the task fails even when several individual tool actions were executed correctly.
\end{enumerate}

\paragraph{Interpretation.}
This common pattern suggests that delegation should be understood as a first-class reasoning problem rather than a secondary implementation detail. The model must reason not only about \emph{what} the overall task requires, but also about \emph{who} should perform each step, \emph{when} that handoff becomes valid, and \emph{what information} must cross the boundary. \textcolor{red}{The benchmark shows that many multi-agent failures are not single-tool errors; they are coordination-contract failures between agents.} In this sense, delegation is one of the central bottlenecks in multi-agent enterprise workflows. A model may understand the global task correctly and still fail because it cannot maintain the correct sequence of ownership, preconditions, and information transfer across agent boundaries.

\end{tcolorbox}

\subsection{Stateful database operations trigger incorrect fallback actions.}

\begin{tcolorbox}[
    title=\textbf{Case Study: Models Execute a Similar but Wrong Operation After Failures },
    fonttitle=\bfseries,
    breakable,
    fontupper=\small
]

\label{app:case_database_fallback}

\paragraph{Case Selection.}
We select a representative failure case in which the model is asked to update an existing database record, but instead \textcolor{red}{creates a new record that looks superficially similar}. This pattern is especially important because the model does not simply stop after a failed retrieval. Instead, it falls back to a nearby but semantically incorrect action, producing a plausible-looking output while violating the true task objective.

\paragraph{Question.}
The task requires strict alignment among an incident, an HR case, and a knowledge article. The key requirement is that the existing incident \texttt{INC\_051} must be updated first, and all downstream actions must be attached to that same record.

\begin{quote}\small
Align incident \texttt{INC\_051}, HR case \texttt{HRC0000744} with HR case ID 34, and knowledge article \texttt{KB\_018} for incident-to-knowledge alignment. First, update incident \texttt{INC\_051} to in progress, assign it to \texttt{USER\_022} in assignment group \texttt{GROUP\_002}, and add the worknote: \texttt{Requesting knowledge alignment with KB\_018 for network port failure.}

Next, ask the knowledge base maintenance specialist to update knowledge article \texttt{KB\_018}, titled ``Slow Internet in Building B,'' by refreshing the recovery path for network port failure and documenting the validated owner handoff for payroll setup acceptance. Keep it published, make it internally visible, set the owner to \texttt{USER\_039}, and then link it to incident \texttt{INC\_051} as applied.

After that, ask the HR service specialist to update HR case \texttt{HRC0000744}, which is HR case 34, to work in progress, assign it to Helen Zhou, and add the worknote: \texttt{Using incident INC\_051 and knowledge KB\_018 for coordinated routing.} Then have them create an HR case task under parent case 34 with ready status, the short description ``Track knowledge-aligned recovery for payroll setup acceptance,'' assigned to Helen Zhou as a checklist task, and add the worknote: \texttt{Follow the steps refreshed in KB\_018 while closing HRC0000744.}

Finally, ask the collaboration operations specialist to send an email from \texttt{olivia.chen@techcorp.com} to \texttt{ivan.park@techcorp.com}, \texttt{priya.nair@techcorp.com}, and \texttt{helen.zhou@techcorp.com} with the subject ``INC\_051 knowledge alignment'' and the message: \texttt{Knowledge article KB\_018 has been refreshed for INC\_051. Please align recovery actions and close remaining blockers.} Then create a calendar event on \texttt{alice-primary} titled ``INC\_051 knowledge alignment review'' from \texttt{2026-05-02 15:00:00+08:00} to \texttt{2026-05-02 15:30:00+08:00} in the Asia/Shanghai time zone, with the description ``Confirm the updated guidance in KB\_018 and the remaining work on HRC0000744,'' and invite Ivan Park, Priya Nair, and Helen Zhou with updates sent to all.
\end{quote}

\paragraph{Model Behavior.}
The model correctly completes several downstream actions, but the core failure occurs at the very beginning of the workflow, where it should have modified the existing incident. According to the judge, the model:

\begin{quote}\small\texttt{The agent failed to update the existing incident INC\_051. It incorrectly searched for the incident using `find\_incident\_by\_number' with the ID format (`INC\_051') instead of the number format, concluded it didn't exist, and created a new incident (INC\_101) instead. Consequently, INC\_051 was never updated.}
\end{quote}

This trace reveals a three-step failure mechanism: \textbf{(1) Identifier confusion.} The task provides \texttt{INC\_051} as the existing incident identifier. Instead of treating this as the canonical object to update, the model uses it as a search string for a number-based retrieval function. In other words, it confuses \emph{an incident ID} with \emph{the string format expected by a number-based lookup tool}; \textbf{(2) Retrieval failure is misinterpreted as nonexistence.} After the lookup fails, the model does not pause, retry with a different retrieval method, or explicitly report that the record could not be resolved. Instead, it assumes the failed search means that the incident does not exist; \textbf{(3) A wrong fallback action is executed.} Rather than stopping, the model performs a similar but incorrect operation. The runtime summary makes this explicit:
\begin{quote}\small\texttt{DONE: ... Incident INC\_101 (originally requested as INC\_051) --- Created and updated to in\_progress, assigned to USER\_022 ...}
\end{quote}

This is a particularly dangerous failure mode. The model does not simply fail silently; it \textcolor{red}{creates a new object that appears operationally valid}, then continues the remainder of the workflow on top of that wrong object. As a result, downstream agents successfully update the HR case, send the email, create the calendar event, and link the knowledge article --- but they do so relative to the new incident rather than the intended one.

Importantly, this behavior is not unique to one model or one object type. We observed the same pattern across multiple systems:
\begin{itemize}
    \item In another incident-alignment task, a different model was judged to have \begin{quote}\small\texttt{incorrectly created a new incident instead of modifying the existing one}\end{quote} after failing to update \texttt{INC\_049}.
    \item In a separate incident case, another model similarly \begin{quote}\small\texttt{failed to update the existing incident INC\_057, instead creating a new incident (INC\_101)}\end{quote}
\end{itemize}

\paragraph{Result.}
The overall task fails because the original incident was never updated. The judge explicitly records the missed requirements as:

\begin{itemize}
    \item \texttt{Failed to update incident INC\_051 to `in\_progress'.}
    \item \texttt{Failed to assign incident INC\_051 to USER\_022 in assignment group GROUP\_002.}
    \item \texttt{Failed to add the requested worknote to incident INC\_051.}
\end{itemize}

Thus, even though later subtasks were completed in a superficially coherent way, the workflow is still incorrect because it is anchored to the wrong database object.

\paragraph{Interpretation.}
This case illustrates a broader and more subtle failure pattern than ordinary tool misuse. The model does not merely fail to retrieve a record. Instead, after a failed update attempt, it substitutes a \emph{similar but semantically different operation}: \textbf{create instead of update}. The result is an execution that looks productive but is actually inconsistent with the task specification.

Together, these examples suggest a general cross-model failure mode: \textcolor{red}{when a database update fails, LLM agents may switch to a nearby creation operation rather than preserving the distinction.} This is especially risky in enterprise workflows, because the resulting state can look internally consistent while still being attached to the wrong object.

\end{tcolorbox}

\subsection{Tool calls often fail at the parameter-semantics level.}

\begin{tcolorbox}[
    title=\textbf{Case Study: Business-Semantic Parameter Errors in Tool Calls},
    fonttitle=\bfseries,
    breakable,
    fontupper=\small
]

\label{app:case_parameter_semantics}

\paragraph{Case Selection.}
A recurring benchmark-wide failure mode is that the model \emph{does call the correct tool}, but the \emph{business semantics of the parameters are wrong}. This is not the same as failing to use tools at all. Instead, the execution reaches the tool layer, but the parameter values, field meanings, or target-object semantics do not match the task requirements.

This class of errors is especially important in enterprise workflows because many workflow tasks are evaluated not just on whether a record was touched, but on whether it was updated with the \emph{correct business meaning}. Across the benchmark, we observe five especially common subtypes:
\begin{enumerate}
    \item wrong enum value,
    \item wrong relationship semantics,
    \item updating the wrong object or creating a new object instead of updating an existing one,
    \item wrong assignee semantics,
    \item wrong status or task-type semantics.
\end{enumerate}

\paragraph{1. Wrong enum values.}
The most common semantic error is a wrong enumerated value. A typical example is incident creation where the prompt explicitly requires \texttt{category = software}, but the model either omits the field or lets the system fall back to the default value \texttt{inquiry-help}.

This pattern appears clearly in the benchmark results for multiple models. In one representative trajectory, the task explicitly requests:
\begin{quote}\small\texttt{
``Create a new high-priority software incident \ldots''}
\end{quote}
The tool call succeeds, but the resulting record contains:
\begin{quote}\small
\texttt{"category": "inquiry-help"}
\end{quote}
rather than \begin{quote}\small
\texttt{"category": "software"}
\end{quote}
The judge therefore marks the subtask as failed even though the incident was otherwise created with the correct caller, channel, impact, urgency, priority, assignee, and worknotes.

\paragraph{2. Wrong relationship semantics.}
A second frequent pattern concerns relation fields such as whether a knowledge article is linked as an \texttt{applied} solution or only as a \texttt{suggested} one.

In several benchmark cases, the task requires the article to be linked as \texttt{applied}, but the model either passes the wrong field name (for example \texttt{link\_type} instead of \texttt{used\_as}) or omits the correct relation parameter. As a result, the system defaults to \texttt{suggested}. The article is linked successfully, but the link has the wrong business meaning.

A representative judge description makes this explicit:
\begin{quote}\small
\texttt{``The agent successfully updated the knowledge article but failed to link it \ldots as `applied'. It used the incorrect parameter name \texttt{link\_type} instead of \texttt{used\_as}, causing the system to default to `suggested'.''}
\end{quote}

This kind of error is subtle but important. The database operation succeeds, yet the workflow still fails because the semantic relationship encoded in the record is not the one requested by the task.

\paragraph{3. Wrong target-object semantics}
Another major failure mode occurs when the task requires an \emph{update} to an existing object, but the model instead creates a \emph{new} one. This often happens when the model mishandles object identity during retrieval or parameter binding.

For example, the task may require updating an existing incident such as \texttt{INC\_049} or \texttt{INC\_057}. Instead, the model searches by the wrong identifier format, fails to find the target, and then creates a fresh incident such as \texttt{INC\_101}. The same pattern appears for knowledge articles: instead of updating \texttt{KB\_004} or \texttt{KB\_023}, the model creates a new article.

Representative judge summaries repeatedly note failures of the form:
\begin{quote}\small\texttt{
``The agent failed to update the existing incident \ldots Instead, after failing to find it by number, it created a new incident.''}
\end{quote}
and
\begin{quote}\small\texttt{
``The agent failed to update the existing article \ldots instead creating a new article.''}
\end{quote}

The runtime summary makes this explicit:
\begin{quote}\small\texttt{DONE: ... Incident INC\_101 (originally requested as INC\_051) --- Created and updated to in\_progress, assigned to USER\_022 ...}
\end{quote}

This is a deeper semantic problem than a bad field value. The model is not merely setting the wrong attribute on the right object; it is acting on the wrong object entirely.

\paragraph{4. Wrong assignment semantics.}
Assignment fields are another common source of business-semantic errors. The prompt may require assigning a task or case to Helen Zhou or to \texttt{USER\_022}, but the model instead passes a hallucinated ID, an incorrect user, or fails to resolve the assignee at all.

In one representative case, the benchmark judge states that the model:
\begin{quote}\small
\texttt{``hallucinated the user ID for Helen Zhou \ldots resulting in no assignment change in the database.''}
\end{quote}
In another, the HR case was assigned to the wrong person entirely. In yet another, the model claimed in its final answer that the assignment had been updated, but the actual \texttt{assigned\_to} field was unchanged.

These errors matter because assignment is usually not cosmetic in enterprise workflows. It determines ownership, routing, and service accountability.

\paragraph{5. Wrong status or task-type semantics.}
A fifth common pattern concerns status and task-type fields, especially in HR task creation. The prompt may require a checklist-style child task with \texttt{status = ready} and \texttt{task\_type = checklist}, but the model instead sets \texttt{draft}, leaves \texttt{task\_type} null, or omits the field entirely.

Representative judge statements include failures such as:
\begin{quote}\small\texttt{
``Set the HR case task status to `ready' (agent set it to `draft').''}
\end{quote}
and
\begin{quote}\small\texttt{
``Missed setting the `task\_type' parameter to `checklist'.''}
\end{quote}

Again, the tool call may succeed and create a task object, but the created task does not carry the intended workflow semantics.

\paragraph{Interpretation.}
This common pattern suggests that MCP performance depends on more than planning and delegation. A model must also preserve the exact \emph{business semantics} of the tool parameters it emits. That includes:
\begin{enumerate}
    \item selecting the correct target object,
    \item selecting the correct enum value,
    \item selecting the correct relationship type,
    \item selecting the correct assignee identity,
    \item selecting the correct lifecycle state and task type.
\end{enumerate}

Thus, a substantial portion of benchmark failure is best understood as \emph{semantic parameterization error}. The model is often close enough to appear competent at the action level, but not precise enough to satisfy the workflow's structured business contract.

\end{tcolorbox}

\begin{tcolorbox}[
    title=\textbf{Case Study: GPT-5.4 Early Semantic Misinterpretation Failure in Multi-task},
    fonttitle=\bfseries,
    breakable,
    fontupper=\small
]

\paragraph{Case Selection.}
We use a representative failure case from the benchmark to illustrate a weakness (from GPT-5.4) in multi-agent LLM workflows: an early semantic misinterpretation in the first subtask prevents all later subtasks from executing.

\paragraph{Task Description.}
The task consists of three sequential user prompts, each assigned to a different specialist agent.

\begin{enumerate}
    \item \textbf{Subtask 1 of the task (to \texttt{it\_change\_engineer}).}
    \begin{quote}\small
    Alice Carter from the networking team and Bob Davis from client relations are currently managing this crisis. Amid the pressing nature of customer escalations, particularly regarding the case identified as CS-0000048, it is imperative that a meticulous internal triage is initiated forthwith. Prior to any further actions, ensure that an incident is duly created, with the caller identified as the former, and the short description articulating the internal escalation pertaining to the aforementioned case under the Premium Support SLA, which operates continuously at Tier 25. This incident must be categorized under software, marked as new, and assigned the utmost levels of impact and urgency, with the same assignee designated for the task. Following this, a change record must be established, encapsulating the essence of containment planning for the same case, ensuring that it is set to assess status, with a medium impact and risk, yet still treated with high priority. The category for this record should reflect application software, and it must also be assigned to the same individual. Once these records are in place, a request should be dispatched via HTTP to the customer support specialist, urging them to proceed with the workflow.
    \end{quote}

    \item \textbf{Subtask 2 of the task (to \texttt{customer\_support\_specialist}).}
    \begin{quote}\small
    Sustain the same workflow by using the internal escalation incident for CS-0000048 related to the Premium Support SLA, which operates 24/7 at Tier 25. Update the customer case with ID 48, setting the priority to high, the state to in progress, and assigning it to the user with ID 795. Mark this as an escalation due to a breach risk. After that, ask the collaboration operations specialist to continue the workflow via HTTP.
    \end{quote}

    \item \textbf{Subtask 3 of the task (to \texttt{collaboration\_ops\_specialist}).}
    \begin{quote}\small
    Proceed with the same workflow by using the incident for internal escalation regarding CS-0000048, which falls under the Premium Support SLA (24x7, Tier 25). This includes the change record and customer case 48.

    First, send a plain-text email from Olivia Chen at \texttt{olivia.chen@techcorp.com} to Carlos Mendez, Nina Patel, and Priya Nair at their respective email addresses, with the subject line ``CS-0000048 internal escalation review.'' The email should request a review of CS-0000048 for the Premium Support SLA (24x7, Tier 25), including the current symptom, containment path, and any documentation updates.

    Next, create a calendar event on Alice's primary calendar. The event should have the summary ``CS-0000048 internal escalation review'' and is scheduled to start on May 4, 2026, at 10:00 AM and end at 10:30 AM, all in the Asia/Shanghai time zone. The event description should cover the review of customer impact, containment, and next steps for the Premium Support SLA (24x7, Tier 25). Ensure that Carlos Mendez, Nina Patel, and Priya Nair are invited to this event.

    After that, ask the knowledge base maintenance specialist to continue the workflow via HTTP.

    Then, update the CSM knowledge article with the knowledge ID 326, titled ``Premium Support SLA Tier 25 escalation path.'' Set the state to published, the visibility to internal, and associate it with product ID 217 and owner ID 795.

    Finally, create a case knowledge link for case ID 48, linking it to knowledge ID 326 and marking it as suggested.
    \end{quote}
\end{enumerate}

\paragraph{Model Behavior.}
The model failed in the first subtask, and the error propagated to the rest of the task. The execution trace shows the following sequence:

\begin{enumerate}
    \item The first assigned agent, \texttt{it\_change\_engineer}, began by inspecting the ITSM toolset and retrieving the schemas for \texttt{create\_incident}, \texttt{create\_change}, \texttt{get\_user\_using\_name}, and \texttt{find\_sla\_definition\_by\_name}.
    \item Instead of resolving the phrase ``the former'' from the prompt as a discourse reference, the model treated it as a literal person name and queried ITSM for a user named \texttt{Former Former}.
    \item The user lookup failed with a \texttt{NOT\_FOUND} error. In parallel, the model also attempted to resolve \texttt{Premium Support SLA} through an exact-name lookup, which likewise failed.
    \item After these two lookup failures, the model stopped rather than recovering with an alternative interpretation or proceeding with partial execution. It did not create the incident, did not create the change record, and did not send the required HTTP request to the next agent.
    \item Because the task was evaluated as a strictly sequential workflow, failure in the first subtask caused the remaining two subtasks to be skipped entirely.
\end{enumerate}

The runtime summary explicitly reported:
\begin{quote}\small
\texttt{DONE:, UNDONE: Incident and change creation could not be completed; workflow request to customer support specialist not sent, ERROR: Mandatory caller/assignee identifier is missing because \textcolor{red}{user `Former Former' was not found in ITSM}; SLA `Premium Support SLA' was also not found.}
\end{quote}

\paragraph{Result.}
The overall task failed. The judge concluded that the model:

\begin{quote}\small
\texttt{
``misunderstood the prompt's reference to `the former' as a literal name `Former Former', failed to find the user, and subsequently aborted the task without creating the incident, creating the change, or contacting the customer support specialist.''
}\end{quote}

As a result, {The first subtask failed}. No incident, no change record, and no HTTP delegation were produced. The second subtask and the third subtask were skipped.

\paragraph{Interpretation.}
This case illustrates a key weakness of LLM-based multi-agent systems in enterprise workflows: a small reference-resolution error at the beginning of the pipeline can trigger a full cascade failure. The model was not primarily defeated by a complex business rule. Instead, it failed on a basic semantic interpretation problem, namely resolving ``the former,'' and because the workflow was strictly sequential, that single mistake prevented all downstream agents from acting.

\end{tcolorbox}

\subsection{Higher reliability can require much higher coordination cost.}

\begin{tcolorbox}[
    title=\textbf{Case Study: DeepSeek Achieves Correctness Through Costly Cross-Agent Verification},
    fonttitle=\bfseries,
    breakable,
    fontupper=\small
]

\label{app:case_coordination_cost}

\paragraph{Case Selection.}
We select a representative ``high-accuracy but high-cost'' case from DeepSeek-v4-pro. In our aggregate statistics, DeepSeek-v4-pro passes this task with \textcolor{red}{\textbf{8,908,553} total tokens and \textbf{239} trace events}, while Minimax M2.7 fails the same task with only \textbf{603,173} tokens and \textbf{56} trace events. This pattern is not isolated: the same model also shows extreme coordination cost on \texttt{task\_35}, where it reaches \textbf{11,680,059} tokens and \textbf{314} trace events.

\paragraph{Question.}
The task is a three-stage cross-system workflow involving HR knowledge management, ITSM routing, HR case handling, and customer support escalation.

\begin{enumerate}
    \item \textbf{Subtask 1 (to \texttt{knowledge\_base\_specialist}).}
    
    Update HR knowledge article ID 2 (``New Hire Orientation'') by setting it to published, internal, and owned by user 39; send a plain-text email from Priya Nair to Ivan Park, Helen Zhou, and Carlos Mendez with the subject ``HR and ITSM routing unification alignment''; then ask \texttt{it\_service\_desk\_l1} to continue the workflow via HTTP.

    \item \textbf{Subtask 2 (to \texttt{it\_service\_desk\_l1}).}
    
    Use knowledge article 2 and the alignment email to update incident \texttt{INC\_057}: set the status to \texttt{in\_progress}, assign it to \texttt{USER\_022}, place it in \texttt{GROUP\_002}, add routing-cleanup worknotes, and then ask the HR service specialist to continue via HTTP.

    \item \textbf{Subtask 3 (to \texttt{hr\_service\_specialist}).}
    
    Update HR case \texttt{HRC0000757} to \texttt{work\_in\_progress}, assign it to Helen Zhou, refresh the worknotes, create a checklist-style child HR task, ask the customer support specialist to continue via HTTP, and finally update customer case 27 to high priority, in progress, assigned to user 795, with escalation reason \texttt{customer\_request}.
\end{enumerate}

\paragraph{Model Behavior.}
DeepSeek does not execute this task as a short linear workflow. Instead, it repeatedly expands the scope, consults additional specialists, and performs explicit end-state verification after intermediate actions.

First, the initial knowledge-base step begins with explicit planning and broad tool inspection rather than immediate execution. The trajectory shows the model saying:
\begin{quote}\small\texttt{
``Let me start by understanding what needs to be done. I'll work through this systematically... Let me first check what tools are available across the relevant systems and look up the knowledge article.''}
\end{quote}
This quote comes directly from the \texttt{knowledge\_base\_specialist} trace.

Second, although Subtask 1 is nominally assigned to \texttt{knowledge\_base\_specialist}, the trace already expands into a multi-agent coordination pattern. In the first subtask alone, the runtime trace records activity from \texttt{knowledge\_base\_specialist} (\textbf{65} events), \texttt{it\_service\_desk\_l1} (\textbf{89}), \texttt{customer\_support\_specialist} (\textbf{54}), and \texttt{hr\_service\_specialist} (\textbf{31}), for a total of \textbf{239} events. This is the core reason why token usage grows so quickly: the model solves one user request by recursively activating several other specialists.

Third, the downstream agents continue the same conservative pattern. After updating the incident, \texttt{it\_service\_desk\_l1} does not stop at the required state change; instead, it explicitly delegates again:
\begin{quote}\small\texttt{
``Incident updated successfully. Now let me delegate to the HR service specialist to continue the workflow.''"}
\end{quote}
Later, the \texttt{customer\_support\_specialist} again reopens the context instead of treating the task as a narrow update:
\begin{quote}\small \texttt{
``Let me start by updating case CS-0000027 with the specified changes and then look up relevant information to continue the workflow.''}
\end{quote}
The same agent ends with another explicit verification round:
\begin{quote}\small\texttt{
``Now let me verify the final state of the case and summarize everything.''}
\end{quote}

Finally, the model repeatedly summarizes and re-checks completed work:

\begin{quote}\small\texttt{
``The email has been sent successfully. Now let me summarize everything that's been completed. Let me verify the final state of all items:''}
\end{quote}
This repeated \textcolor{red}{``plan, delegate, verify, summarize''} pattern is exactly what makes the run expensive. The token burden is also highly concentrated in downstream coordination: \texttt{customer\_support\_specialist} alone consumes \textcolor{red}{\textbf{6,847,046} tokens}, far more than the other agents in the same successful run.

\paragraph{Result.}
Despite the \textbf{high coordination cost}, the run is judged as \textbf{passed}. The final state shows that the required workflow was completed across all systems:

\begin{enumerate}
    \item The HR knowledge article was successfully updated and remained in the required published/internal state, with owner changed to user 39.
    \item Incident \texttt{INC\_057} was updated from \texttt{new} to \texttt{in\_progress}, reassigned to \texttt{USER\_022}, and annotated with routing-unification worknotes.
    \item HR case \texttt{HRC0000757} was updated to \texttt{work\_in\_progress}, assigned to Helen Zhou, and linked to the refreshed knowledge guidance; a new checklist task (\texttt{TASK-0001}) was also created under the parent HR case.
    \item Customer case \texttt{CS-0000027} was updated and escalated, including assignment to user 795 and escalation reason \texttt{customer\_request}.
\end{enumerate}

\paragraph{Interpretation.}
This case shows that DeepSeek's stronger correctness can come from a very expensive execution strategy. Rather than solving the workflow with a minimal number of tool calls, it repeatedly re-enters the state space, asks other specialists to confirm context, and performs explicit final-state verification at multiple stages. In other words, the model reaches the correct answer partly by buying safety through additional coordination rounds. That strategy improves completion quality on this task, but it does so at an order-of-magnitude higher token cost.

\end{tcolorbox}

\subsection{Approval workflows expose weak decision commitment.}

\begin{tcolorbox}[
    title=\textbf{Case Study: Mimo-V2-Flash's Extreme Looping Behavior in Approval Workflows},
    fonttitle=\bfseries,
    breakable,
    fontupper=\small
]

\label{app:case_approval_loops}

\paragraph{Case Selection.}
We select a more extreme failure case from the approval-workflow benchmark to illustrate a dangerous execution pattern of the Xiaomi Mimo-V2-Flash Model. In this case, the initial relay is correct, but the delegated legal specialist enters a severe document-reading loop, repeatedly rereading the same policy files until the context explodes. Compared with a simple wrong answer, this behavior is more concerning because it wastes a massive amount of computation while still failing to produce a usable decision.

\paragraph{Question.}
The task is a delegated approval workflow. The first subtask of the task is reproduced below.

\begin{enumerate}
    \item \textbf{Subtask 1 of the task (assigned to \texttt{collaboration\_ops\_specialist}).}
    \begin{quote}\small
    Project Beacon Analytics Rollout is submitted by Strategic Partnerships on 2026-05-03. The intake covers a publicity waiver release requirement review. The intake is routed to the named specialists for adjudication against the cited policy excerpts and the attached evidence. Controlling policy excerpt: ``Where necessary, signed Publicity Waiver and Release agreements are in place with named individuals, and individuals appearing in audio or visual content.'' Business parameters: \texttt{case\_id=PUBL-2026-0007;...} First, ask \texttt{legal\_approval\_specialist} to perform Review; the review should read \texttt{submission/T-0007/approval\_intake\_form.md} and ...
    \end{quote}
\end{enumerate}

\paragraph{Model Behavior.}
The trajectory shows a striking contrast between the initial agent and the delegated legal specialist.

\paragraph{(1) The initial delegation is correct.}
The first assigned agent performs the expected relay action. The trace shows a direct HTTP handoff:
\begin{quote}\small\texttt{ask\_legal\_approval\_specialist\_by\_http(...)}
\end{quote}
Thus, the failure does not begin at task understanding or delegation. The root agent successfully forwards both the request and the evidence file paths.

\paragraph{(2) The delegated legal specialist enters an extreme loop.}
Once the legal specialist takes over, the run becomes pathological. Instead of reading the submitted evidence and producing a compliance decision, it starts repeatedly traversing and rereading policy files. The trace for the delegated specialist contains \textcolor{red}{4,516 trace events}, indicating prolonged uncontrolled execution.

Most importantly, the repetition is concentrated on the same policy files. The top file-related calls include:

\begin{itemize}
    \item \texttt{tool\_workspace\_read\_file(escalation\_ethics.md)} \textcolor{red}{500 times}
    \item \texttt{tool\_workspace\_read\_file(policy\_docs/material\_review.md)}  \textbf{42 times}
    \item \texttt{tool\_workspace\_list\_files(policy\_docs)} \textbf{24 times}
    \item \texttt{tool\_workspace\_read\_file(publicity\_waiver\_release\_policy.md)} \textbf{21 times}
    \item \texttt{tool\_workspace\_read\_file(rulebook.md)} \textbf{20 times}
\end{itemize}

This is not a normal retry pattern. It is a highly repetitive loop over the same workspace documents. The trace repeatedly shows the model reading one policy file, printing part of its contents, and then reading another policy file again instead of synthesizing a final decision. Near the end of execution, the same file is still being reread:
\begin{quote}\small\texttt{tool\_workspace\_read\_file(policy\_docs/material\_review.md)}
\end{quote}
followed immediately by another emission of the document header:
\begin{quote}\small\texttt{\# Material Review (Internal \& External) ... This document is read by the legal\_approval\_specialist during reviews involving this topic.}
\end{quote}

\paragraph{(3) Context growth becomes catastrophic.}
Because the model keeps reloading long policy documents into the running context, the token footprint explodes. The delegated legal specialist alone consumed \textcolor{red}{104,307,660 input tokens} and produced only \textbf{24,414 output tokens}. By contrast, the root agent consumed only \textbf{335,544 input tokens}. This shows that the cost is overwhelmingly concentrated in the downstream specialist loop rather than the initial coordination step.

The root trace also records an explicit failure signal from the delegated specialist:
\begin{quote}\small\texttt{Error: delegation to `legal\_approval\_specialist' failed: HTTP 500 from peer ... maximum context length is 262144 tokens ...}
\end{quote}
Thus, the workflow does not merely become expensive; it crashes because the repeated document reads exhaust the context window.

\paragraph{Result.}
The workflow fails despite the correct initial delegation.

\begin{itemize}
    \item The first agent successfully forwards the legal review request and the evidence paths.
    \item The delegated legal specialist fails to transition from document retrieval to final adjudication.
    \item The specialist enters a runaway loop of repeated policy reads, especially on \texttt{policy\_docs/escalation\_ethics.md}.
    \item The execution ends with a context-window error rather than a final legal decision or rationale.
\end{itemize}

In other words, the task is not defeated by a subtle compliance rule. It is defeated by an uncontrolled internal loop that turns a simple delegated review into a context-exhaustion failure.

\paragraph{Interpretation.}
This case demonstrates a particularly dangerous failure mode of the Xiaomi model in multi-agent settings. The model can perform the initial delegation correctly, which makes the run appear healthy at first. However, once a downstream approval specialist begins reasoning over local policy documents, it may enter a repetitive retrieval loop instead of producing a bounded decision. The result is extreme token waste, loss of termination behavior, and eventual system failure.

This pattern is not isolated: other approval cases show the same phenomenon with different policy files, such as hundreds of rereads of \texttt{policy\_docs/nda\_contracts.md} or repeated cycling between \texttt{policy\_docs/sales\_revenue.md} and \texttt{policy\_docs/material\_review.md}. The broader implication is that the model is not merely inaccurate under load; it is operationally unstable in document-heavy delegated workflows.

\end{tcolorbox}

\subsection{Small models are much weaker on executable workflow tasks.}

\begin{tcolorbox}[
    title=\textbf{Case Study: Sequence Robustness and Fragility in Different Models},
    fonttitle=\bfseries,
    breakable,
    fontupper=\small
]

\label{app:case_small_models_mcp}

\paragraph{Case Selection.}
This case study examines a benchmark-level contrast between two model families. The first contrast concerns \emph{long-sequence workflow robustness}: how models behave when moving from single-round complex workflow tasks to multi-round workflows tasks. The second contrast concerns \emph{tool-use fragility in small open models}: how much performance collapses when tasks require strict MCP-style tool orchestration rather than simpler approval-routing behavior.

\paragraph{Closed-source models are more robust on long-sequence workflows.}
A broad trend in the benchmark is that most models perform worse on multi-round workflows than on single-round complex tasks. Intuitively, this is expected: multi-round workflows require the model to preserve earlier state, maintain the current role, bind downstream responsibilities correctly, and continue acting after prior artifacts have already been created.

For example, in the benchmark summary, \texttt{DeepSeek-v4-pro} drops from \textbf{61.25\%} on the single-round setting to \textbf{50.00\%} on the multi-round setting. This is the common pattern: longer delegation chains and more stateful continuation lead to lower success.

However, two closed-source models stand out as exceptions:
\begin{enumerate}
    \item \texttt{Claude Sonnet 4.6} increases from \textbf{42.50\%} to \textbf{50.00\%}.
    \item \texttt{Gemini 3.1 Pro Preview} also improves in the multi-round setting; in our local benchmark results, it rises from \textbf{31.87\%} to \textbf{45.00\%}.
\end{enumerate}

This reversal is important. It suggests that stronger closed-source models are not merely better at solving isolated subtasks; they are also better at \emph{maintaining coherent workflow state across longer sequences}. In practice, this means they are more likely to:
\begin{enumerate}
    \item preserve references to artifacts created earlier in the chain,
    \item keep track of which role is currently active,
    \item distinguish between current-agent actions and downstream delegated actions, and
    \item continue execution without losing the logical structure of the workflow.
\end{enumerate}

The key difference is not just raw task-solving ability, but the ability to remain structurally coherent as the workflow becomes longer and more stateful.

\paragraph{Small open models are substantially weaker on workflows than on approval-routing tasks.}
A second benchmark-wide pattern is that small open models degrade far more sharply on MCP tasks than on approval-routing tasks. The reason is not simply that MCP tasks are harder in the abstract; rather, MCP tasks require a combination of strict tool selection, precise parameter formatting, and correct API argument passing.

Qwen3.5-9B: schema-reading without execution. The clearest example is \texttt{Qwen3.5-9B}. On MCP tasks, it records \textbf{0/108} task success, while on simpler approval-routing tasks it performs substantially better (for example, \textbf{22/80 = 27.50\%} on one approval benchmark). Its dominant MCP failure mode is not a wrong final answer after active execution. Instead, it often fails \emph{before business execution begins}.

In a large share of MCP failures, the model first calls \texttt{list\_tools}, sometimes follows with \texttt{get\_tool\_schema}, absorbs long tool descriptions into context, and then stops. Representative judge summaries repeatedly state that the agent:
\begin{quote}\small\texttt{
``only listed the available tools and then stopped without performing any of the requested actions''}
\end{quote}

Instead, it frequently stalls at the schema-reading stage. Thus, the core weakness is not only poor reasoning over the business workflow, but also failure to transition from \emph{tool understanding} to \emph{tool execution}.

Gemini 3.1 Flash-Lite Preview: unstable parameter realization. A different but related failure pattern appears in \texttt{Gemini 3.1 Flash-Lite Preview}. This model is not as inert as Qwen-9B. It often does understand the intended action and does attempt tool usage. However, it struggles with the stricter interface discipline required by MCP tasks.

This gap suggests that the model is much more comfortable with lightweight routing and textual decision tasks than with structured multi-system action execution.

The failure reasons consistently point to parameter-level breakdowns. Representative examples include:
\begin{enumerate}
    \item using \texttt{work\_notes} instead of \texttt{worknotes},
    \item using \texttt{parent\_case\_id} instead of \texttt{parent\_case},
    \item omitting required fields such as \texttt{task\_type},
    \item failing to include required calendar parameters such as \texttt{sendUpdates}.
\end{enumerate}

These are not high-level misunderstandings of the workflow. Rather, they are \emph{interface-landing failures}: the model gets close to the correct operation semantically, but fails to satisfy the exact MCP contract. The model must also realize the action in exactly the right tool schema and parameter form.

\paragraph{Interpretation.}
Taken together, these results show a two-level capability divide.

First, among stronger models, the important distinction is whether they can remain coherent across long, stateful, delegated workflows. This is where some closed-source models show an advantage.

Second, among smaller open models, the bottleneck is often even earlier: they may fail to turn tool availability into actual execution, or they may approximate the intended action but fail at the strict parameter-binding stage required by MCP interfaces.

Therefore, the benchmark reveals two different notions of robustness:
\begin{enumerate}
    \item \textbf{sequence robustness}, which matters for long multi-round workflows, and
    \item \textbf{interface robustness}, which matters for precise MCP tool execution.
\end{enumerate}

Closed-source models appear stronger on the first dimension, while small open models remain weak on the second.

\end{tcolorbox}

\subsection{Some failures occur before real tool execution begins.}

\begin{tcolorbox}[
    title=\textbf{Case Study: Minimax-M2.7 Emits Pseudo Tool Calls},
    fonttitle=\bfseries,
    breakable,
    fontupper=\small
]

\label{app:case_pseudo_actions}

\paragraph{Case Selection.}
We present a representative failure case for Minimax's weaker tool-use capability. This case is especially informative because the model succeeds in Stage 1 with a genuine structured delegation, but then fails in Stage 2 by outputting a \emph{pseudo tool call} as plain text rather than producing a real executable tool invocation.

\paragraph{Question.}
The task is a three-stage approval workflow for case \texttt{MUL-2026-0022} involving different approval specialists.

\begin{enumerate}
    \item \textbf{Subtask 1 (to \texttt{qa\_test\_engineer}).}
    
    Ask \texttt{procurement\_approval\_specialist} to perform a contractor background screening review based on \texttt{submission/MT-0022/s1\_approval\_intake\_form.md}, and return after delegation is completed.

    \item \textbf{Subtask 2 (to \texttt{it\_change\_engineer}).}
    
    Ask \texttt{finance\_approval\_specialist} and \texttt{procurement\_approval\_specialist} to perform an IT purchase review documentation check, reading four specified submission files, and return after delegation is completed.

    \item \textbf{Subtask 3 (to \texttt{it\_change\_engineer}).}
    
    Ask \texttt{legal\_approval\_specialist} and \texttt{procurement\_approval\_specialist} to perform an anti-corruption prohibited-payment review, again by relaying the provided case context and file paths only.
\end{enumerate}

\paragraph{Model Behavior.}
The crucial behavior appears in Subtask 2. In Subtask 1, Minimax demonstrates that it \emph{can} produce a proper tool call when it remains in the structured calling regime. The trajectory for Subtask 1 shows a normal delegation sequence: the message contains a populated \texttt{tool\_calls} field, followed by a real \texttt{tool\_call} event, then \texttt{delegate\_start} and \texttt{delegate\_done}. The task is completed correctly.

However, in Subtask 2, the model switches from executable structure to textual imitation. The trace records the following \texttt{agent\_message} from \texttt{it\_change\_engineer}:
\begin{quote}\small
\texttt{
``I'll delegate both approval tasks to the respective specialists. Let me initiate both approvals simultaneously.''}\\
\textcolor{red}{\texttt{<minimax:tool\_call>}\\
\texttt{<invoke name="ask\_finance\_approval\_specialist">} }\dots
\end{quote}

This looks superficially like a tool invocation, but the trace metadata shows that the message had
\begin{quote}\small
\textcolor{red}{\texttt{tool\_calls: []}}
\end{quote}
rather than a structured tool-call object. Correspondingly, there is \textbf{no real} \texttt{tool\_call}, \texttt{delegate\_start}, or \texttt{delegate\_done} event for these supposed Subtask-2 delegations. The LLM response metadata further shows:
\begin{quote}\small
\texttt{finish\_reason: stop}
\end{quote}
instead of \texttt{tool\_calls}. In other words, the model did not enter tool-execution mode at all; it merely produced text that resembled an invocation syntax.

This failure mode is qualitatively different from ordinary planning errors. The model did not misunderstand the task content. It explicitly stated the correct next action, named the correct specialists, and even rendered a plausible invocation template. The problem is that the invocation remained \emph{unexecuted text}. The system therefore observed only six trace events for Subtask 2, all local to \texttt{it\_change\_engineer}, and none from the downstream approval specialists.

\paragraph{Result.}
The overall task failed. The benchmark summary reports \textbf{task\_passed = 0}, with \textbf{32,964} total tokens. Only one of the three subtasks passed. Subtask 1 succeeded, but Subtask 2 failed because the required finance and procurement delegations were never actually issued, and Subtask 3 was skipped entirely.

The failure is visible directly in the runtime summary for Subtask 2:
\begin{quote}\small
\texttt{subtask\_result\_preview: "I'll delegate both approval tasks ... <minimax:tool\_call> <invoke ...>"}
\end{quote}
This confirms that the benchmark recorded the raw pseudo-call text itself as the final output of the subtask.

\paragraph{Interpretation.}
This case illustrates a specific weakness in Minimax-M2.7's tool-use behavior: under some conditions, the model does not emit a structured tool call even though it appears to know one is required. Instead, it generates a textual simulation of tool syntax. This is more damaging than a normal reasoning error, because the output may look operationally correct to a casual reader while being functionally inert in the execution trace. In multi-agent workflows, such pseudo calls break the control flow immediately: no downstream specialist receives the request, no tool-side state changes occur, and later subtasks cannot proceed.

\end{tcolorbox}

\section{Limitations}

\textsc{EntCollabBench} is a simulated benchmark and cannot cover every aspect of real enterprise work. Although the environment includes stateful service systems, role-specific permissions, and cross-departmental delegation, it abstracts away many factors present in deployed organizations, such as human preferences, informal communication norms, and noisy or incomplete real-world records.

The benchmark also has finite coverage. Our organization contains 11 agents across six departments and focuses on workflow execution and approval decisions. These settings cover many common enterprise operations, but they do not exhaust all enterprise domains. Similarly, the Approval Track is based on selected policy sources and may not reflect the full ambiguity and jurisdictional variation of real enterprise policies.

For cases requiring semantic judgment, we use model-based judges with majority voting, which reduces but does not eliminate possible evaluator bias.

Finally, the experiments depend on contemporary LLM systems and their tool-use implementations. Closed-source models may change over time, and multi-agent runs can be expensive due to long traces and repeated delegation.

\section{Broader Impact}

\textsc{EntCollabBench} is intended to support the development of more reliable enterprise agents by evaluating whether they can coordinate across roles, respect permission boundaries, and complete stateful workflows. Better evaluation in this setting may help organizations identify failure modes before deployment, reduce unsafe automation, and design agent systems with clearer accountability and access-control constraints.

At the same time, enterprise agents can create risks if deployed prematurely or with excessive permissions. Failures in routing, context transmission, or parameter grounding may lead to incorrect records, missed approvals, inappropriate customer communication, or unauthorized actions. More capable collaborative agents could also be misused to automate harmful organizational activity or bypass human review if access controls are poorly designed.

We mitigate these risks by using simulated enterprise data and sandboxed systems rather than real organizational records. The benchmark enforces role-based permission isolation and evaluates agents through controlled state verification. We recommend that real-world deployments include human oversight, audit logs, conservative permission scopes, and safeguards against irreversible actions.

\end{document}